\documentclass{pasj01}

\makeatletter
\let\leftroot\relax
\let\uproot\relax

\makeatother
\usepackage{amsmath} 
\usepackage{graphicx}

\usepackage{CJKutf8}
\newcommand{\namecn}[1]{\begin{CJK*}{UTF8}{gbsn}({#1})\end{CJK*}}
\newcommand{\namejp}[1]{\begin{CJK*}{UTF8}{min}({#1})\end{CJK*}}

\usepackage{array}
\usepackage{multirow}
\usepackage{tabularx}  
\usepackage{graphicx}
\usepackage{lastpage}

\usepackage{chngcntr} 

\usepackage[hyphens]{url}
\usepackage{microtype}

\usepackage[switch,mathlines]{lineno}  
\usepackage[normalem]{ulem}  

\usepackage{flafter} 
\usepackage{fix-cm} 

\Received{$\langle$19 Aug 2025$\rangle$}
\Accepted{$\langle$6 Jan 2026$\rangle$}
\Published{$\langle$publication date$\rangle$}

\begin{document}

\title{ 
Unraveling Year-Long Radial Velocity Variations in Red Clump Region -- I: Comprehensive analysis of a K0 Giant star, 2 Draconis.
}

\author{
Udomlerd {Srisuchinwong} \namecn{徐奕昌} \altaffilmark{1,2}
}
\email{udomlerds@bao.ac.cn}
\author{   
Jianzhao {Zhou} \namecn{周建召} \altaffilmark{3,4}
}
\author{      
Huan-Yu {Teng} \namecn{滕环宇} \altaffilmark{1,5,6}
}
\email{hyteng@bao.ac.cn}
\author{      
Guang-Yao {Xiao} \namecn{肖光耀} \altaffilmark{7,1}
}
\author{      
Bun'ei {Sato} \namejp{佐藤文衛} \altaffilmark{5}
}
\author{      
Takuya {Takarada} \namejp{宝田拓也} \altaffilmark{8,9}
}
\author{    
Masashi {Omiya} \namejp{大宮正士}
\altaffilmark{8,9}
}
\author{      
Hiroki {Harakawa} \namejp{原川紘季} \altaffilmark{10}
}
\author{      
Eiji {Kambe} \namejp{神戸栄治}  \altaffilmark{10}
}
\author{      
Hideyuki {Izumiura} \namejp{泉浦秀行}  \altaffilmark{11}
}
%
\author{
Michitoshi {Yoshida} \namejp{吉田道利} \altaffilmark{9}
}
\author{      
Yoichi {Itoh} \namejp{伊藤洋一} \altaffilmark{12}
}
\author{      
Hiroyasu {Ando} \namejp{安藤裕康} \altaffilmark{9}
}
\author{      
Eiichiro {Kokubo} \namejp{小久保英一郎} \altaffilmark{9,13}
}
\author{      
Marc {Hon} \altaffilmark{14}
}
\author{      
Yujuan {Liu} \namecn{刘玉娟} \altaffilmark{1}
}
\author{      
Fei {Zhao}  \namecn{赵斐} \altaffilmark{1}
}
\author{      
Wei {Wang} \namecn{王炜} \altaffilmark{1}
}
\author{      
Meng {Zhai} \namecn{翟萌}  \altaffilmark{1}
}
\author{      
Shaolan {Bi} \namecn{毕少兰}  \altaffilmark{3,4}
}
\author{      
Gang {Zhao} \namecn{赵刚} \altaffilmark{1}
}
\email{gzhao@nao.cas.cn}

\altaffiltext{1}{
National Astronomical Observatories, Chinese Academy of Sciences, Beijing 100012, China
}
\altaffiltext{2}{
School of Astronomy and Space Science, University of Chinese Academy of Sciences, Beijing 100049, China
}
\altaffiltext{3}{
Institute for Frontiers in Astronomy and Astrophysics, Beijing Normal University, Beijing 102206, China
}
\altaffiltext{4}{
Department of Astronomy, Beijing Normal University, Beijing 100875, China 
}
\altaffiltext{5}{
Department of Earth and Planetary Sciences, Institute of Science Tokyo, 2-12-1 Ookayama, Meguro-ku, Tokyo 152-8551, Japan 
}
\altaffiltext{6}{
Korea Astronomy and Space Science Institute, 776 Daedeok-daero, Yuseong-gu, Daejeon 34055, Republic of Korea
}
\altaffiltext{7}{
Tsung-Dao Lee Institute, Shanghai Jiao Tong University, 1 Lisuo Road, Shanghai, 201210, China
}
\altaffiltext{8}{
Astrobiology Center, National Institutes of Natural Sciences, 2-21-1 Osawa, Mitaka, Tokyo 181-8588, Japan
}
\altaffiltext{9}{
National Astronomical Observatory of Japan, National Institutes of Natural Sciences, 2-21-1 Osawa, Mitaka, Tokyo 181-8588, Japan
}
\altaffiltext{10}{
Subaru Telescope, National Astronomical Observatory of Japan, National Institutes of Natural Sciences, 650 North A’ohoku Pl., Hilo, HI, 96720, USA
}
\altaffiltext{11}{
Okayama Branch, Subaru Telescope, National Astronomical Observatory of Japan, National Institutes of Natural Sciences, Kamogata, Asakuchi, Okayama 719-0232, Japan
}
\altaffiltext{12}{
Nishi-Harima Astronomical Observatory, Center for Astronomy, University of Hyogo, 407-2, Nishigaichi, Sayo, Hyogo 679-5313, Japan
}
\altaffiltext{13}{
The Graduate University for Advanced Studies (SOKENDAI), 2-21-1 Osawa, Mitaka, Tokyo, 181-8588, Japan
}
\altaffiltext{14}{
Kavli Institute for Astrophysics and Space Research, Massachusetts Institute of Technology, Cambridge, MA 02139, USA
}

\KeyWords{
stars: individual: HD100696 ---
planets: exoplanets (498) --- 
techniques: radial velocity (1332)
}

\maketitle

\begin{abstract}
Slow-rotating evolved stars frequently exhibit radial velocity (RV) variations on annual timescales, complicated by instrumental systematics and aliasing in the one-year regime.  
Here we investigate the origin of the near-yearly periodicity in 2 Dra, a star located in the red-clump region, assessing possible causes between stellar activity, instrumental profile (IP) effects, sampling alias, and planetary companions. 
We applied two independent approaches: (1) constraining diagnostic signals and performing a correlation analysis ($r$) between period-confined signals, and (2) evaluating phase stability by partitioning Keplerian fits. These methods enabled us to examine the physical connections and phase coherence among stellar activity indicators, RV measurements, and IP diagnostics. Our analysis suggests a stellar rotation period of $\simeq270\text{--}320$\,d for 2~Dra. The 340-d RV signal does not appear to originate from stellar activity in this chromospherically quiet star ($|r| \lesssim 0.33$), nor from instrumental systematics near the annual period ($|r| \lesssim 0.1$). This conclusion is supported by contrasting phase behavior: the RV and stellar activity phases remain stable, whereas the IP phases do not. We therefore propose that the 340-d variation likely arises from either small-amplitude intrinsic variability or a tentative gas giant companion with potential weak activity-induced modulation. The case of 2~Dra provides a framework for distinguishing the origins of $\sim$1-yr RV variations in other evolved stars.
\end{abstract}


\section{Introduction} 
\label{sec: intro}
The detection of long-period exoplanets via precise radial velocity (RV) measurements is often falsely identified  by spurious signals with periods near one year. This critical periodicity region is frequently contaminated by overlapping sources of variability, particularly the sampling effect of yearly observations. Long-term instrumental drift and stellar activity can also mask or mimic the reflex motion induced by a planet. For instance, \citet{dumusque2015characterization} and \citet{ribas2018candidate} identified artificial yearly signals in HARPS data resulting from instrumental distortions linked to Earth's annual orbit. A similar case was reported by \citet{anglada2014two}, who attributed a 340-day signal from Kapteyn's star to seasonal measurement uncertainties or stellar rotation. However, \citet{robertson2015stellar} later demonstrated that the star has a 143-day rotation period, revealing the putative 48-day super-Earth signal to be merely a rotational harmonic. Collectively, these cases demonstrated that annual aliases and instrumental systematics can introduce spurious periodicities if left uncorrected, particularly near one-year periods.

Several methods have been developed to mitigate these yearly variability issues. These approaches are as follows. (1) Alias cleaning: Deconvolving the window function (WF) identifies and removes aliases \citep{desort2009extrasolar}. (2) Data correction: Removing spectral lines affected by yearly systematics or subtracting a one-year sinusoidal fit directly from the RV data \citep{dumusque2015characterization}. (3) Sparse signal recovery: Using the $l_1$ periodogram recovers true signals in multi-planet systems, even when the highest peak in a Lomb–Scargle periodogram is a spurious alias \citep{hara2017radial}. (4) Pipeline processing: Applying the \texttt{YARARA} pipeline involves principal component analysis on line-by-line RVs to reduce stellar, instrumental, and one-year signals, thereby revealing low-amplitude, long-period planets \citep{cretignier2023yarara}. Despite these promising advances, a significant challenge remains when stellar, instrumental, and sampling-related peaks are closely spaced or blended within the same one-year region, particularly when the spurious signals are weak.

Long-period variability in evolved stars, particularly K-giants, commonly exhibits periods of $\sim200\text{-}800$ days. Notable examples include $\gamma$ Dra (702 d, K5III; \cite{hatzes2018radial}), $\alpha$ Tau A (629 d, K5III; \cite{hatzes2015long}), and 42~Dra (479 d, K1.5III; \cite{dollinger2009planetary}). These variations were initially identified as planet candidates in RV searches, but they were later refuted when their non-Keplerian nature became apparent through phase and amplitude shifts linked to stellar sources \citep{hatzes2018radial, reichert2019precise, heeren2021precise, hatzes2025no}. However, current models have not yet fully explained all such variations in evolved stars.

This presents two key observational challenges. First, long-period giant planets orbiting evolved stars frequently exhibit orbital periods in the hundreds-of-days range \citep{sato2005radial, dollinger2009planetary}. This overlaps with the periodicities of intrinsic stellar variations in K-giants. Second, the RV amplitudes of candidate planets are often comparable to stellar jitters (typically tens to hundreds of $\mathrm{m}\,\mathrm{s^{-1}}$), as demonstrated in $\epsilon$ Cyg \citep{heeren2021precise}. Thus, amplitude alone is not an unreliable discriminator and makes it difficult to distinguish between planetary and stellar origins of the observed signals. In particular, when the variability occurs near the one-year region, it is still debatable whether these stellar variations mainly account for significant RV variations in K-giant stars. Disentangling their origins becomes especially challenging.

To address these challenges, our strategy is to systematically exclude perturbation effects (e.g., stellar activity, instrumental systematics, sampling) from close RV variations to validate genuine planetary signals. We further assess variability sources by testing for consistency in period, amplitude, and phase, as true periodic signals should maintain stability across these parameters, unlike non-periodic stellar or instrumental effects. Our goal is to establish a comprehensive framework for distinguishing and diagnosing the sources of RV modulations, specifically addressing the 1-year degeneracy problem in slow-rotating evolved stars. We demonstrate this approach using the slow-rotating giant 2~Dra (HD 100696, TIC 142827466), which exhibits $\sim$1-year degenerated variations in our HIDES (HIgh Dispersion Echelle Spectrograph) data.

We organize our paper into 7 sections. Section \ref{sec: observation} details the observations of spectroscopy with RV measurement and \textit{TESS} photometry. Section \ref{sec: stellar_model} represents the stellar models of 2~Dra regarding the stellar parameter derivation and asteroseismology. Section \ref{sec: periodogram_analysis} explains the periodogram analysis and our correlation analysis model. Section \ref{sec: determine_nature_signal} diagnoses the variability nature of RV, stellar signals of line activity, and instrumental signals of HIDES instruments. Section \ref{sec: kep_orbit_fit} shows Kepler orbit fitting and planet models. Section \ref{sec: discussion} discusses the year-long variations of 2~Dra RV signals in each variation source.

\enlargethispage{\baselineskip}

\section{Observations}\label{sec: observation}
\subsection{Spectroscopy}
\label{subsec: spectroscopy}
Our spectral observation of 2~Dra was acquired between Feb 2002 and Apr 2021 by the 188-cm telescope at Okayama Astrophysical Observatory with the HIDES \citep{izumiura1999hides} to target the long-period giant planet \citep{sato2005radial}. We collected stellar spectra for RV measurement with monthly cadences using three distinct instrumental configurations, chronologically as follows.  (\textit{i}) a slit configuration (hereafter, HIDES-S) observed during 2002-2016 \citep{kambe2002development}. (\textit{ii}) $1^{\rm st}$ fiber-link upgrade (hereafter, HIDES-F1) measured during 2011-2017 \citep{kambe2013fiber}, and (\textit{iii}) $2^{\rm{nd}}$ fiber-link upgrade (hereafter, HIDES-F2) observed during 2019-2021. Data taken with each configuration have independent RV zero-point offsets, which are treated separately in the analysis.

For slit-mode observations, HIDES-S was set up by  $200\text{-}\mu m$ slit width ($0.\!''76$) with the spectral resolution $R \sim 67000$ around 3.3 pixel sampling with 20-min exposure time and signal-to-noise ratio ($S/N$) over 150. For fiber-mode observations, HIDES-F1 and HIDES-F2 were installed by $1.\!''05$ sliced image-width with the resolution $R \sim 55000$ about 3.8 pixel sampling, enhancing $S/N$ ratio $>200$ with exposure time to 10 min. Later, the high-efficiency fiber-link system and optical instruments were reconfigured in 2018 to improve the optical stability of instruments (reducing thermal fluctuation and vibration) in a temperature-controlled room \citep{teng2023revisiting}. 

The \'{e}chelle data reduction (bias subtraction, flat-fielding, scattered-light subtraction, and spectrum extraction) was implemented via IRAF\footnote{ IRAF is distributed by the National Optical Astronomy Observatories, which is operated by the Association of Universities for Research in Astronomy, Inc. under a cooperative agreement with the National Science Foundation, USA} package in the conventional manner. Precise RV measurement analysis in each observed mode was derived by following \citet{sato2002development} and \citet{butler1996attaining}'s methods by using $\rm{I_2}$ cell for calibrating the wavelength reference (5000–5800 \AA) and tracking instrumental shift. We adopted the modeled spectrum via the stellar template spectra superposed by high-resolution $\rm{I_2}$ spectra (i.e., $\rm{I_2}$-superposed stellar spectra), where both $\rm{I_2}$ and stellar line spectra were convoluted with the spectrograph's instrument spectra, i.e., instrument profile (IP). The accuracy of RV measurements was determined by the performance of the spectrograph's IP modeling \citep{sato2002development}. We obtained the stellar template spectra as follows: \textit{i})  for HIDES-S via the deconvolution of the pure stellar spectrum with IP determined from $\rm{I_2}$-superposed B-type star spectrum, and \textit{ii}) for HIDES-F1 and -F2 via the direct high-resolution observations ($R \sim 100000$) without additional lines from $\rm{I_2}$ cell. Our finalized RV values with errors were determined from the average of velocity measurements over hundred sections of the dissected spectra ($\sim3$\AA) in the individual instruments. For instrumental stability, we referred the reader to read further in Appendix 2 of \citet{teng2022regular}. RV data of 2~Dra by the 1.88-m telescope was provided through \texttt{GitHub} \footnote{https://github.com/okayamapsp/okayamapsp.github.io}.

Upgrading HIDES instruments in 2007 from a single CCD to a three-CCD mosaic broadened the three wavelength regions to 3700-7500 \AA \ as new reference lines. This enabled the concurrent measurements of precise RV, stellar line profiles, and chromospheric activity in Ca \textsc{ii} H\&K lines (at 3968.47 \AA, 3933 \AA). Regardless, as a result of the high scattered light in wavelengths of 3700-4000 \AA \ by the extreme aperture overlaps in fiber-modes, Ca \textsc{ii} H line spectra of both HIDES-F1 and -F2 were unable to be removed by IRAF, thereby necessarily excluded.

\subsection{\textit{TESS} Photometry}
\label{subsec: photometry_tess}
\textit{TESS} (Transiting Exoplanet Survey Satellite), a space-based telescope launched in 2018 \citep{ricker2015transiting}, provided high-precision photometry for characterizing planets and stellar variability. Imaging of 2~Dra was observed in several sectors from 2019 to 2024 (Sector 14, 21, 41, 47, 48, 74, 75) chronologically, spanning a $\sim$27 observing period in each sector. This satellite provided full-framed images in different minute cadences: 2-min (short cadence), 10-min (intermediate cadence), and 30-min exposures (long cadence). The observation of 2~Dra was performed under multi-cadence coverage with data-reduction pipelines, primarily SPOC pipeline (Science Processing Operations Center) for short cadence data \citep{jenkins2016tess}.

To preserve the coherent light curve analysis with high-fidelity, we selected the 2-min cadence \textit{TESS} light curve data with zero quality flags from SPOC pipeline as a basis for asteroseismology. The gaps in the light curves between sectors were present as a result of the interruption of data transmission back to Earth.

\section{Stellar models} 
\label{sec: stellar_model}
\begin{table}[!htbp]
\caption{Stellar parameters of 2~Dra}
\begin{tabularx}{0.52\textwidth}{lc}  
\hline\hline
Parameters & Values \& Sources \\
\hline\hline
Spec. type  & K0III$^{\rho}$\\ 
Evolutionary phase & $\text{RGB}^{\triangle}$ \\
Parallax $\pi$ [mas] & $13.49 \pm 0.51^{h}$ \\
Distance $d$ [pc] & ${80.67^{+0.41}_{-0.40}}^{\dagger, b}$ \\
Visual magnitude 
$V$ & $5.305 \pm 0.002^{h}$ \\
Color index $B - V$ & 0.974$^{h}$ \\
Effective temperature $T_\text{eff}$ [K] & ${4842.15^{+17.83}_{-10.44}}^{\dagger,g}$ \\
& $4833 \pm 18^{*}$ \\
Surface gravity $\log g_{*}$ [cgs] & ${2.415^{+0.007}_{-0.007}}^{\dagger,g}$ \\
& $2.32 \pm 0.06^{*}$ \\
Metallicity $[\text{Fe}/\text{H}]$ [dex] & ${-0.313^{+0.047}_{-0.005}}^{\dagger,g}$ \\
& $-0.33 \pm 0.04^{*}$ \\
Stellar mass $M_{*}$ [$M_\odot$] & ${1.124^{+0.031}_{-0.036}}^{\dagger,g}$ \\
& ${1.021^{+0.664}_{-0.664}}^{\dagger,b}$ \\
Stellar radius $R_{*}$ [$R_\odot$] & ${10.888^{+0.081}_{-0.085}}^{\dagger,g}$ \\
& ${10.638^{+0.170}_{-0.167}}^{\dagger,b}$ \\
Luminosity $L_{*}$ [$L_\odot$] & ${58.74^{+1.09}_{-0.94}}^{\dagger,g}$ \\
& ${55.43^{+1.58}_{-1.51}}^{\dagger,b}$ \\
Proj. rotational velocity $v\sin i$ [$\mathrm{km}\,\mathrm{s}^{-1}$] & ${1.68 \pm 0.11}^{*}$ \\
Max stellar rotation period  $\dfrac{P_{\rm{rot},*}}{\sin i}\ [\rm{d}]$$^{(\psi)}$ & ${328.19^{+23.57}_{-20.67}}^{\dagger,g,*}$ \\
& ${320.65^{+23.47}_{-20.66}}^{\dagger,b,*}$ \\
Max oscillation power’s freq. $\nu_{\rm{max}}$ [$\mu \rm{Hz}$] & ${35.98 \pm 0.55}^{\triangle}$\\
Large freq. separation 
$\Delta\nu$  [$\mu \rm{Hz}$] & ${4.23 \pm 0.06}^{\triangle}$ \\
Age [Gyr] & ${5.38^{+0.58}_{-0.55}}^{\dagger,g}$ \\
Prob. of RGB phase & ${0.964}^{\triangle}$ \\
\hline
\end{tabularx}
\vspace{1mm}
{\footnotesize \raggedright 
Note: $^{*}$ = Spectroscopy \citep{takeda2008stellar}. \\
$^{\dagger}$ = Evolutionary Track (This work). \\
$^{\triangle}$ = Asteroseismology (This work). \\
$^{g}$ = Grid model (main default value, if compared). \\
$^{b}$ = Bolometric correction.
\\
$^{h}$ = Hipparcos \citep{esa1997vizier}. 
$^{\rho}$ = \cite{roman1952spectra}. \\
$^{(\psi)}$ = Use $P_{\text{rot}, *}/\sin i = 2\pi R_{*}/(v\sin i).$ \\
\par}
\label{tab: stellar_param}
\end{table}
\begin{figure}
    \centering
    \includegraphics[width=1.\linewidth]{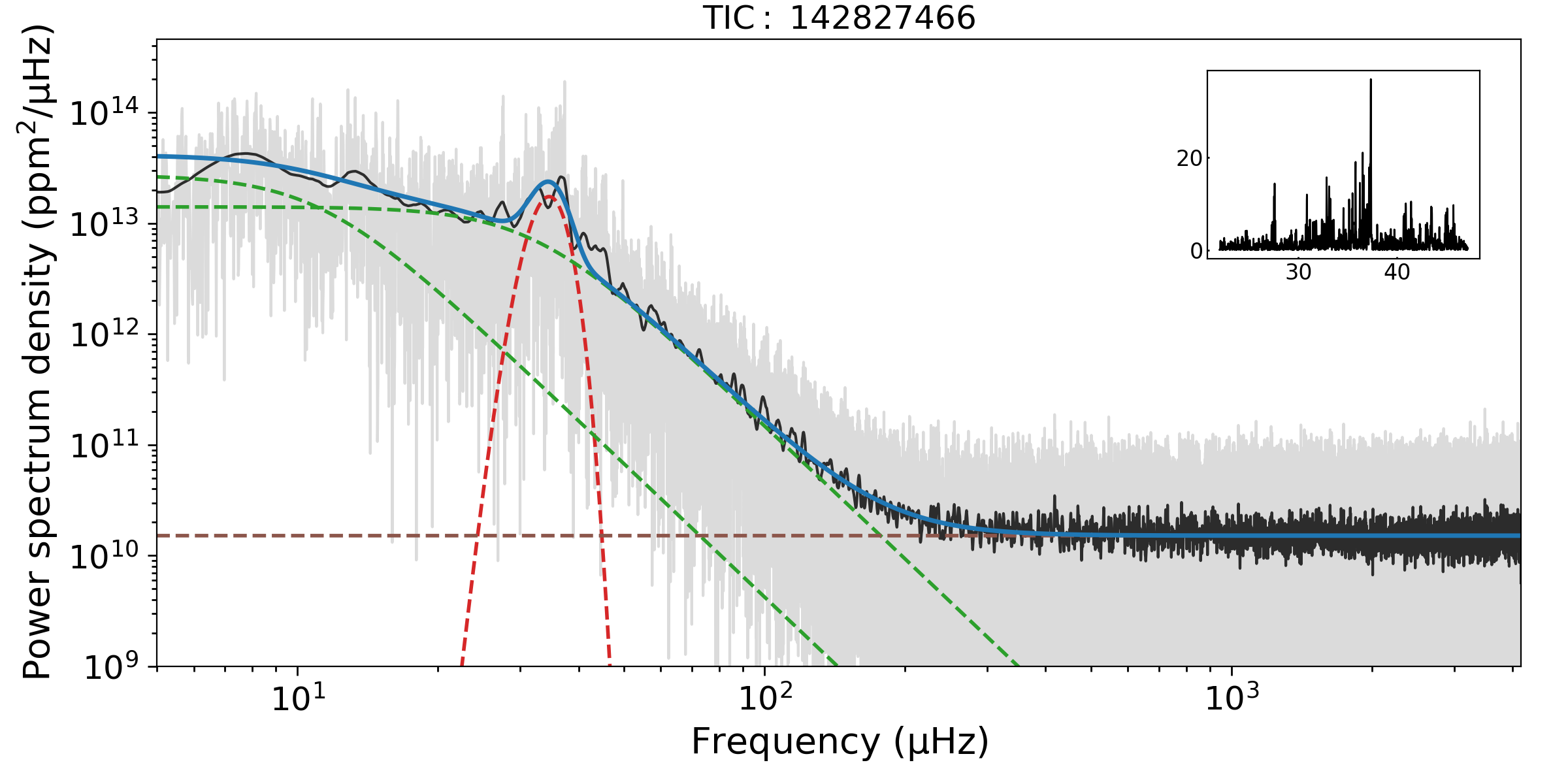}
    \caption{
    Power spectrum density (PSD) diagrams of \textit{TESS} lightcurve of 2~Dra. 
    \textit{Main panel:} The original PSD is shown in \textit{gray curve}, while the smoothed PSD is a \textit{black curve}. The oscillation hump is represented in the Gaussian \textit{red curve},   the two granulation terms are present in the two \textit{green curves}, the white noise is a \textit{brown line}, and the sum of \texttt{MCMC} fit components is for the \textit{blue curve}.
    \textit{Top right panel} is the non-smoothed background (granulation and white noise)-removed  PSD (in unit of dex in power). 
    }
    \label{fig: Power Spectrum}
\end{figure}
\begin{figure}
    \centering
    \includegraphics[width=1.\linewidth]{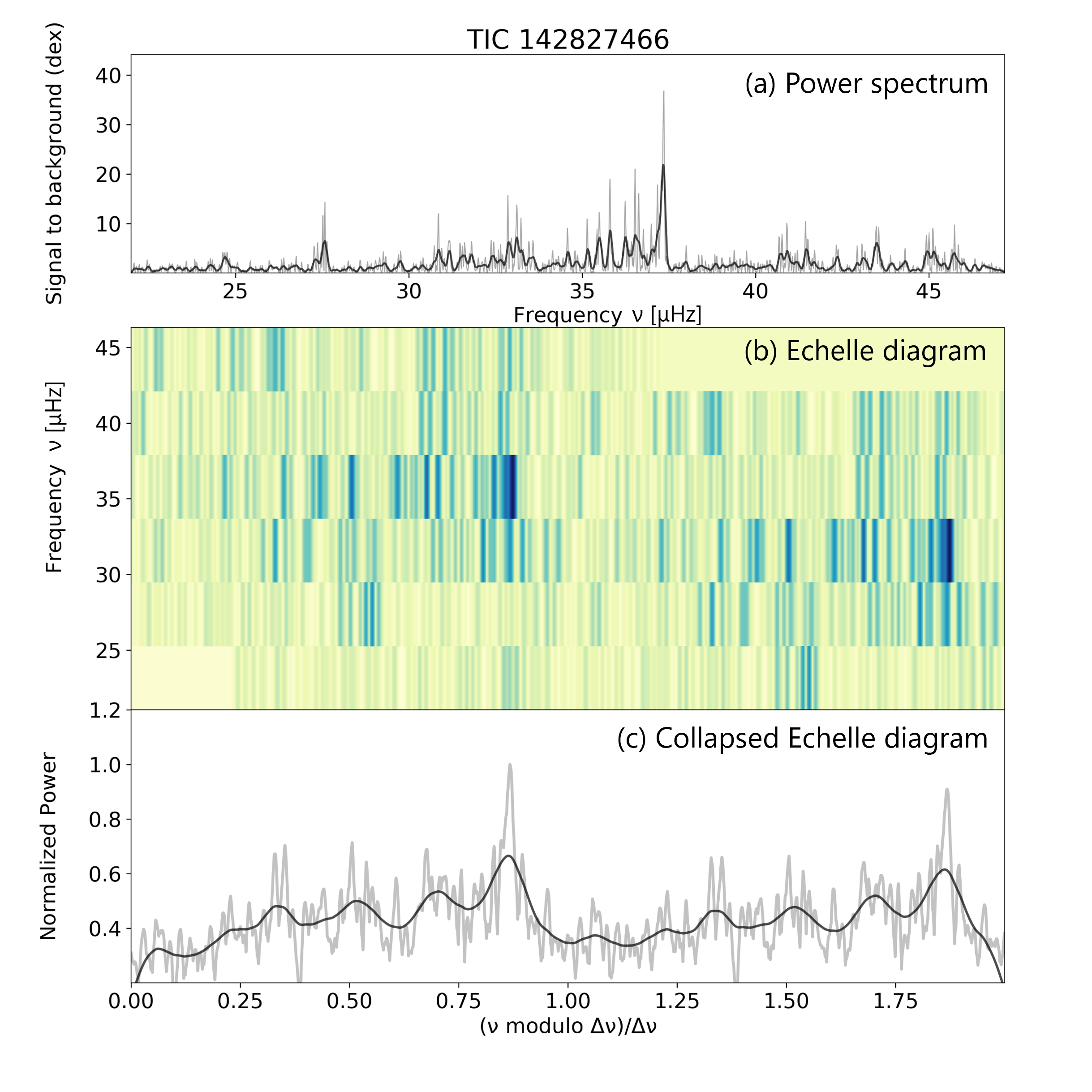}
    \caption{
    Asteroseismic diagnostic plots of 2~Dra. 
    \textit{Top panel}: the smoothed PSD with the oscillation peak at $\nu_{\text{max}}=35.98 \pm 0.55\,\mu\rm{Hz}$ and background (same line colors as main Figure \ref{fig: Power Spectrum}). 
    \textit{Middle panel}: \'{E}chelle diagram (ED) showing the radial oscillation modes with the highest power at $\nu_{\text{max}}$ (\textit{darkgreen}).
    \textit{Bottom panel}: Collapsed \'echelle diagram showing the highest peak with its uncertainty at $\Delta\nu = 4.23 \pm 0.06\,\mu\rm{Hz}$ (following \textit{darkgreen} grid in ED), as the frequency width of radial modes.
    }
    \label{fig: echelle_diagram}
\end{figure}
\begin{figure}[!ht]
    \centering
    \includegraphics[width=1.\linewidth]{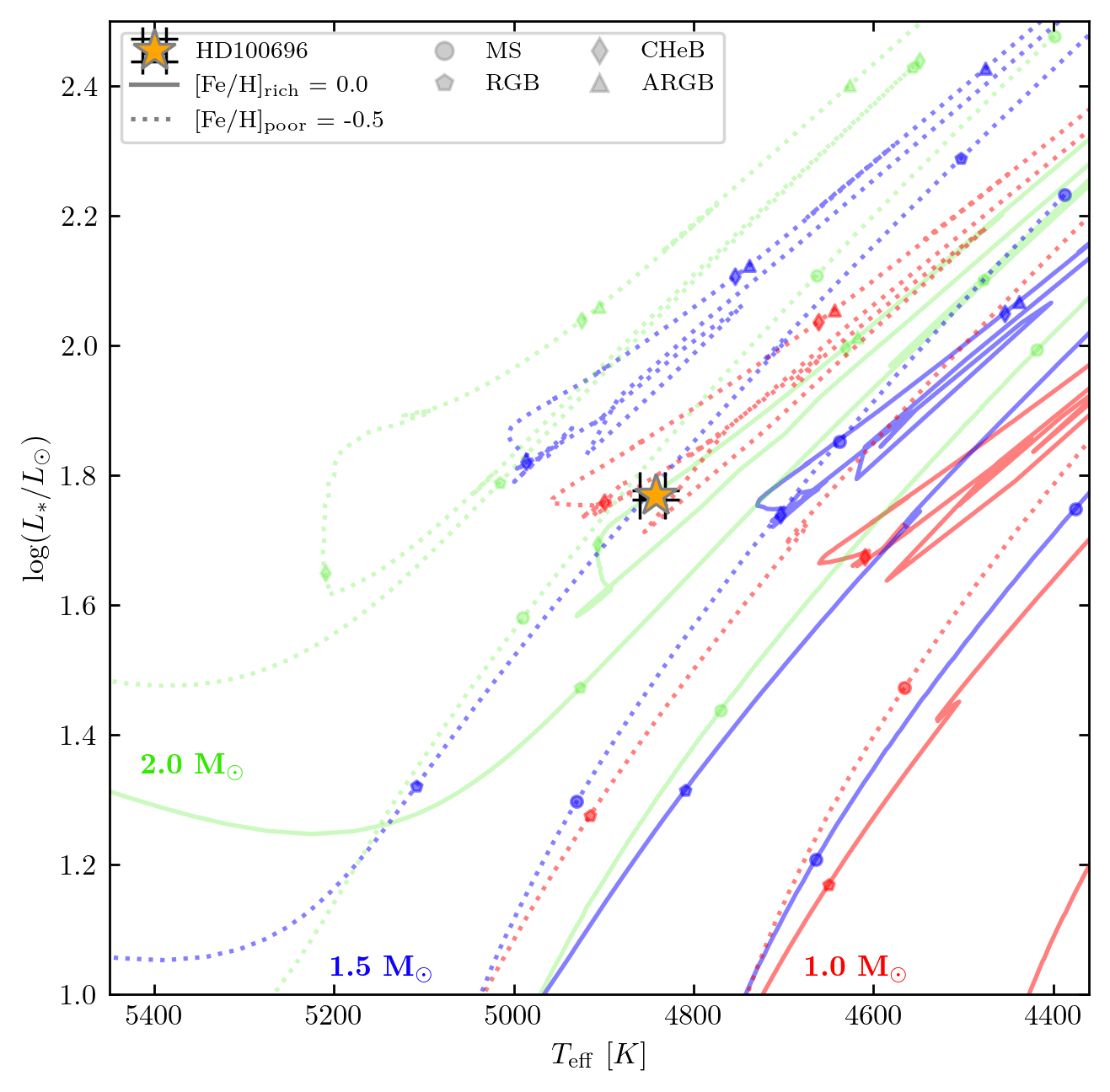}
    \caption{Evolutionary tracks (HR diagram) of the evolved intermediate-mass giant stars around solar-masses ($1-2M_{\odot}$) in different stages: main sequence (MS), red giant branch (RGB), core-He burning (CHeB), and asymptotic red giant branch (ARGB), between the margin of metal-poor $(\text{[Fe/H]} = -0.5$; dashed line) and metal-rich content $(\text{[Fe/H]} = 0.0$; solid line). 2~Dra is located with a white star with errorbar. 
    }
    \label{fig: HR_diagram}
\end{figure}

Our model determined the stellar parameters of 2~Dra jointly via asteroseismology from \textit{TESS} photometry and stellar track (isochrone) classification in Hertzsprung-Russell (HR) diagram. 

From spectroscopy, the atmospheric parameters with errors of 2~Dra (i.e., effective temperature $T_{\text{eff}}$, surface gravity $\log g_{*}$, and metallicity $\text{[Fe/H]}$) were derived in $\rm{I}_{2}$-free stellar spectra via the equivalent width measurement of Fe \textsc{i} and Fe \textsc{ii} lines by \citet{takeda2008stellar} in relation to the dispersion of Fe abundance. This produced the spectral $T_{\text{eff}} = 4833 \pm 18 \text{ K}$, $\log g_{*} = 2.32 \pm 0.06 \text{ cgs}$, and  $\text{[Fe/H]} = -0.33 \pm 0.04 \text{ dex}$, respectively. The projected rotational velocity $(v \sin i)$ was measured through spectral line broadening analysis connecting to rotational and turbulent parameters \citep{takeda2008stellar}.

From photometry, the 2~Dra's asteroseismic parameters (i.e., max radial-oscillation power’s frequency $\nu_{\text{max}}$, {large frequency separation} $\Delta\nu$) with errors were derived from \textit{TESS} light curves following the {similar} methodology as described in \citet{xiao2024two}. The Discrete Fourier Transform (DFT) approach was applied to \textit{TESS} photometric data for {generating} the power spectrum density ({PSD;} \cite{VanderPlas2018}). 
{
Figure \ref{fig: Power Spectrum} displays the resulting PSD computed by classical FT, illustrating the excess power from the oscillation signal and the granulation background; the best-fit Gaussian envelope used to determine $\nu_{\text{max}}$ is measured, and the locations of the radial-mode ridge and the derived $\Delta\nu$ are indicated in \'{E}chelle diagram (Figure \ref{fig: echelle_diagram}).} 
The {PSD} was fitted using the Maximum Likelihood Estimate (MLE) method. {This fit} was finalized with Markov-Chain Monte-Carlo (\texttt{MCMC}) {to refine parameters with} a Gaussian envelope, three background Harvey components, and white noise, following \texttt{SYD} pipeline \citep{huber2009automated, Chontos2022pySYD}. The autocorrelation function (ACF) method was employed to measure $\Delta\nu$ values. {$\Delta\nu$ uncertainties} were obtained by subjecting the PSD to 500 perturbations, employing a $\chi^2$ distribution with two degrees of freedom. We estimated the signal-to-background ratio at  $\nu_{\text{max}}$ to be $\sim20$ (Figure \ref{fig: echelle_diagram}, panel a), revealing the oscillation peak. The oscillation excess is also assured by the collapsed-\'echelle peak and similarly consistent across multiple radial orders. To assess the robustness of $\Delta\nu$, our \'{e}chelle diagram (Figure \ref{fig: echelle_diagram}), folded at the measured $\Delta\nu$ of the PSD, reveals a clear band of radial modes at the expected frequency width $\Delta\nu$ with regularly-spaced oscillation frequencies, ensuring the large separation for a solar-like oscillator. Notably, red giants often exhibit mixed $p$–$g$ modes and broader linewidths, deviating mode patterns from simple radial oscillation \citep{Beck2011fast}. Still, the radial ridge (e.g., $l=0$) here is relatively stable to measure $\Delta\nu$ - sufficient to characterize the solar-like $p$-modes. (\cite{Tassoul1980asymptotic}; further mixed-mode discussion: \citet{Beck2011fast}). The evolutionary phase was determined by a deep learning classifier \citep{hon2017deep, hon2018deep, hon2022hd}, which {used} the frequency distribution of oscillation modes within a star’s collapsed \'{e}chelle diagram. Consequently, we determined the star as a {Red Giant Branch (RGB)} with a probability of {96.4}\%, having $\nu_{\text{max}}={35.98 \pm 0.55}\,\mu\rm{Hz}$ and $ \Delta\nu = {4.23 \pm 0.06}\,\mu\rm{Hz}$. Notably, the period spacing $\Delta\Pi$ of $l=1$ oscillation mode is not derivable due to the insufficient signal-to-noise of the PSD. Thus, we could not determine the  evolutionary phase of 2~Dra by $\Delta\Pi$.

In stellar track classification, we modeled the isochrone fits via the Bayesian approach with \texttt{isoclassify} package \citep{huber2017asteroseismology, berger2020gaia}. {We incorporated}  the spectroscopy (spectral $T_{\text{eff}}$), photometry (Tycho $B,V$; Gaia $G,G_\mathrm{BP},G_\mathrm{RP}$; SDSS $r$; 2MASS $J,H,K_{s}$ magnitudes), parallax ($\pi$), and asteroseismology parameters ($\nu_{\text{max}}, \Delta\nu$), to derive the stellar parameters. {These parameters were determined} by {two methods}. (\textit{i}) {One is} the grid interpolation of evolutionary tracks in HR diagrams (i.e., grid model method). (\textit{ii}) {Another is} the bolometric correction with dust extinction map ({by} \texttt{mwdust} package; \cite{bovy2016galactic}) via empircal relationship (i.e., direct method).

In the grid approach, the integrated model with isochrones derived the posterior $T_{\text{eff}}=4842.15^{+17.83}_{-10.44} \text{ K}$, $\log g_{*}=2.415^{+0.007}_{-0.007} \text{ cgs}$, [Fe/H] $=-0.313^{+0.047}_{-0.005} \text{ dex}$, stellar mass $M_{*}= 1.12^{+0.03}_{-0.04} \ M_{\odot}$, radius $R_{*}= 10.89^{+0.08}_{-0.09} \ R_{\odot}$, luminosity $L_{*}= 58.74^{+1.09}_{-0.94} \ L_{\odot}$, $\text{Age} = 5.38^{+0.58}_{-0.55} \text{ Gyr},$ and interstellar extinction $A_{V} = 0.02^{+0.04}_{-0.04} $ considering in $V$-band map \citep{green20193d}. In the direct approach, we derived the bolometric correction  ($BC = M_{\text{bol}} - M_{V}$) by interpolating the parameters of $(T_{\text{eff}}, \log g_{*}, \text{[Fe/H]}, A_{V})$ with {Monte-Carlo} sampling in the MESA Isochrones $\&$ Stellar Tracks (MIST) grid \citep{paxton2010modules, choi2016mesa}. This provides the direct posterior mass $M_{*}=1.02^{+0.66}_{-0.66} \ M_{\odot}$, radius $R_{*}=10.64^{+0.17}_{-0.17} \ R_{\odot}$, luminosity $L_{*}=55.43^{+1.58}_{- 1.51} \ L_{\odot}$, distance $d= 80.67^{+0.41}_{-0.40} \text{ pc}$, $V$-band absolute magnitude $M_{V}= 0.657^{+0.030}_{-0.031}$, and bolometric magnitude $M_{\text{bol}}=0.381^{+0.031}_{-0.030}$ via the empirical relations. Incorporating $v\sin i$,  we derived the stellar rotation period $P_{\text{rot},*}\simeq320\text{ - }328 \text{ day}$ in both modes, assuming the edge-on view, representing a slow-rotating star. Figure \ref{fig: HR_diagram} indicates the evolutionary tracks from MIST and the evolved star location of 2~Dra in the HR diagram with different stages and iron abundances.

The stellar parameters are derived in Table \ref{tab: stellar_param}. In our analysis, the parameters in the grid evolution track model served as inputs for the RV model. Particularly, the stellar mass $M_{*}$ {was used} as a key indicator to determine the minimal planet mass in RV analysis in Section \ref{sec: kep_orbit_fit}.

\section{Periodogram Analysis} 
\label{sec: periodogram_analysis}
\begin{figure*}[ht!]
    \centering
    \includegraphics[width=0.95\linewidth]{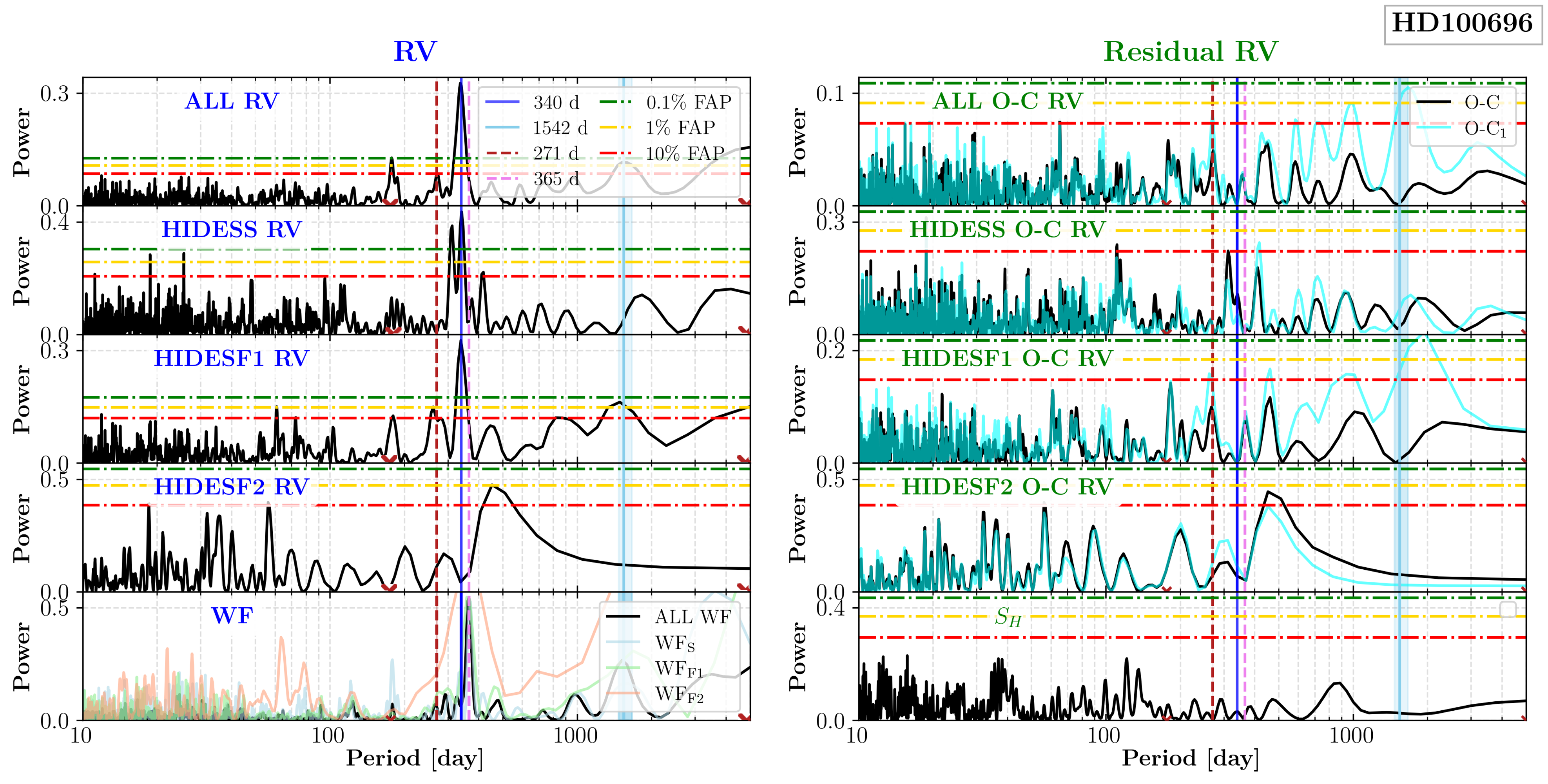}
    \caption{GLS-Ps of 2 Dra's RV variations (\textit{left panels}) with RV residuals (\textit{right panels}), WF (\textit{bottom left panel}), and $S_{H}$ (\textit{bottom right panel}). 
    In RV residual panel, $\rm{O-C}_{1}$ is the $\rm{1^{st}}$ RV residual after 340-d RV peak removed, and subsequently, $\rm{O-C}$ is the last RV residual after 1540-d RV peak removed. 
    Three FAP levels are represented by green (0.1\%), yellow (1\%), and red (10\%) dot-dash lines in each indicator, respectively. Vertical skyblue band is a hypothetical outer period errorbar. The alias signals ($\times$ marker) are indicated at the zero baseline ($P_{\rm{alias}} [\text{yr}] = 1 / |1/P_{\text{peak}} \pm 1|$; \cite{dumusque2012earth}).
    }
    \label{fig: GLS_p_rv_2Dra}
\end{figure*}
\begin{figure*}[ht!]
    \centering
    \includegraphics[width=0.9\linewidth]{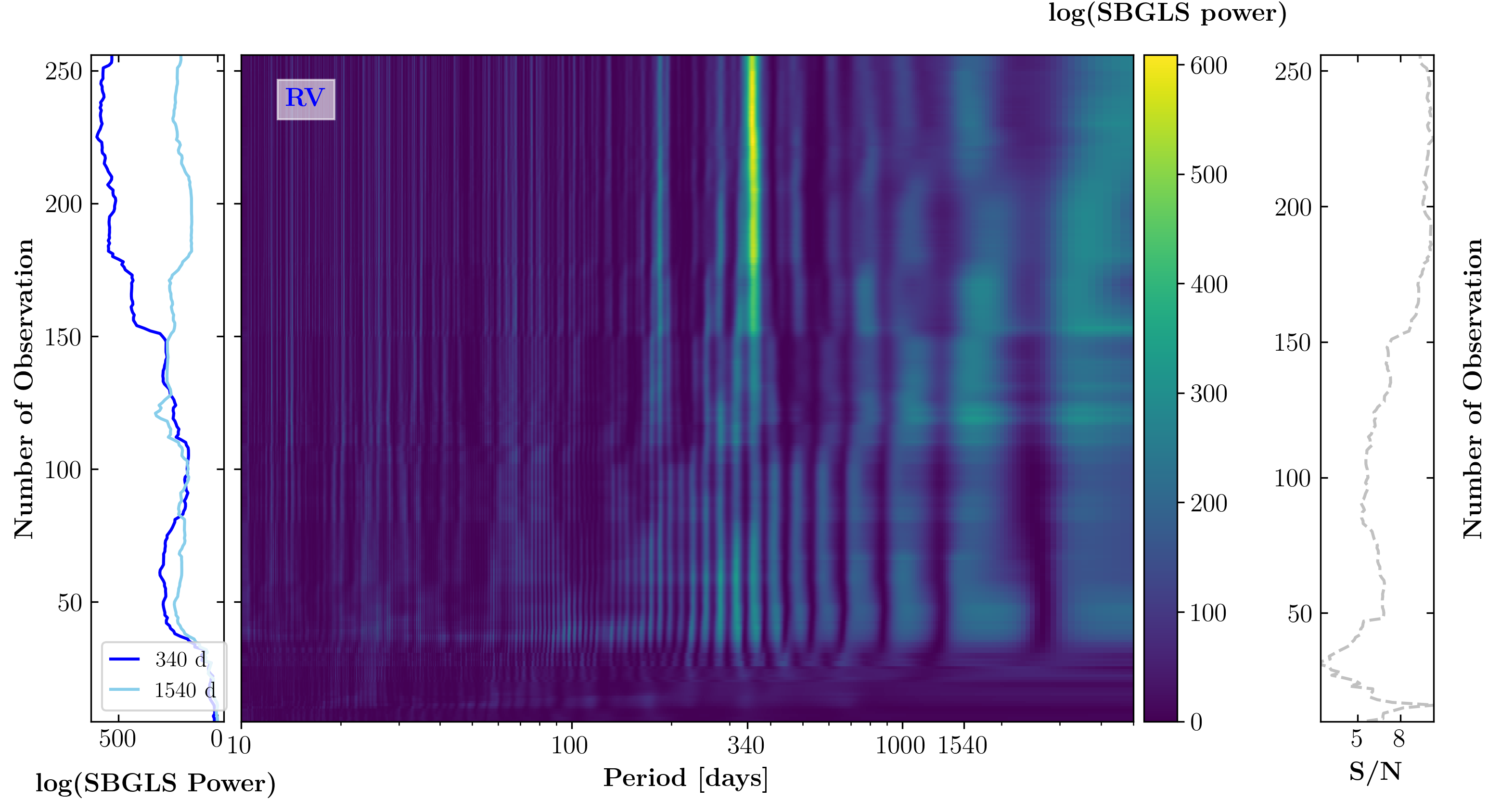}
    \caption{SBGLS-P of RV variations of 2~Dra. 
    Central colormap signifies SBGLS-P power (i.e., the significance probability) across the signal periods over observation time. Left vertical plot is the projected SBGLS-P probability at the significant periods (e.g., 340-d and 1540-d RV, respectively). Right vertical plot is the max-likelihood signal-to-noise ratio over the number of observational data given input jitters $s_{\rm{inst}}$ computed by \texttt{radvel} in each instrument.
    }
    \label{fig: SBGLS_rv}
\end{figure*}
To identify significant periodicities in {RV} time series, we performed two period analyses. {First is a} Generalized Lomb-Scargle Periodogram (hereafter, GLS-P; \cite{zechmeister2009generalised}) with false alarm probability (FAP) via \citet{baluev2008assessing} method. {Second is a} Stacked Bayesian GLS-P (hereafter, SBGLS-P). {This is} to validate if periodic signals are coherently significant over time by increasing data {via} maximized Bayesian likelihood and signal-to-noise {$(S/N)$} ratio, following \citet{mortier2017stacked} method. We selected significant signals using two criteria: peaks with $\leq 0.1\%\ \text{FAP}$ in the GLS-P (excluding aliases and harmonics) and significant upward probability in the SBGLS-P analysis. GLS-P plots for RVs of 2~Dra are present in Figure \ref{fig: GLS_p_rv_2Dra}. SBGLS-P plot for RV diagnostics is shown in Figure \ref{fig: SBGLS_rv}, while GLS-P and SBGLS-P plots for stellar and instrumental indicators are in Sections \ref{subsubsec: activity_periodogram_analysis} and \ref{subsubsec: inst_profile_analysis}, respectively.
\begin{figure}[!ht]
    \centering
    \includegraphics[width=1\linewidth]{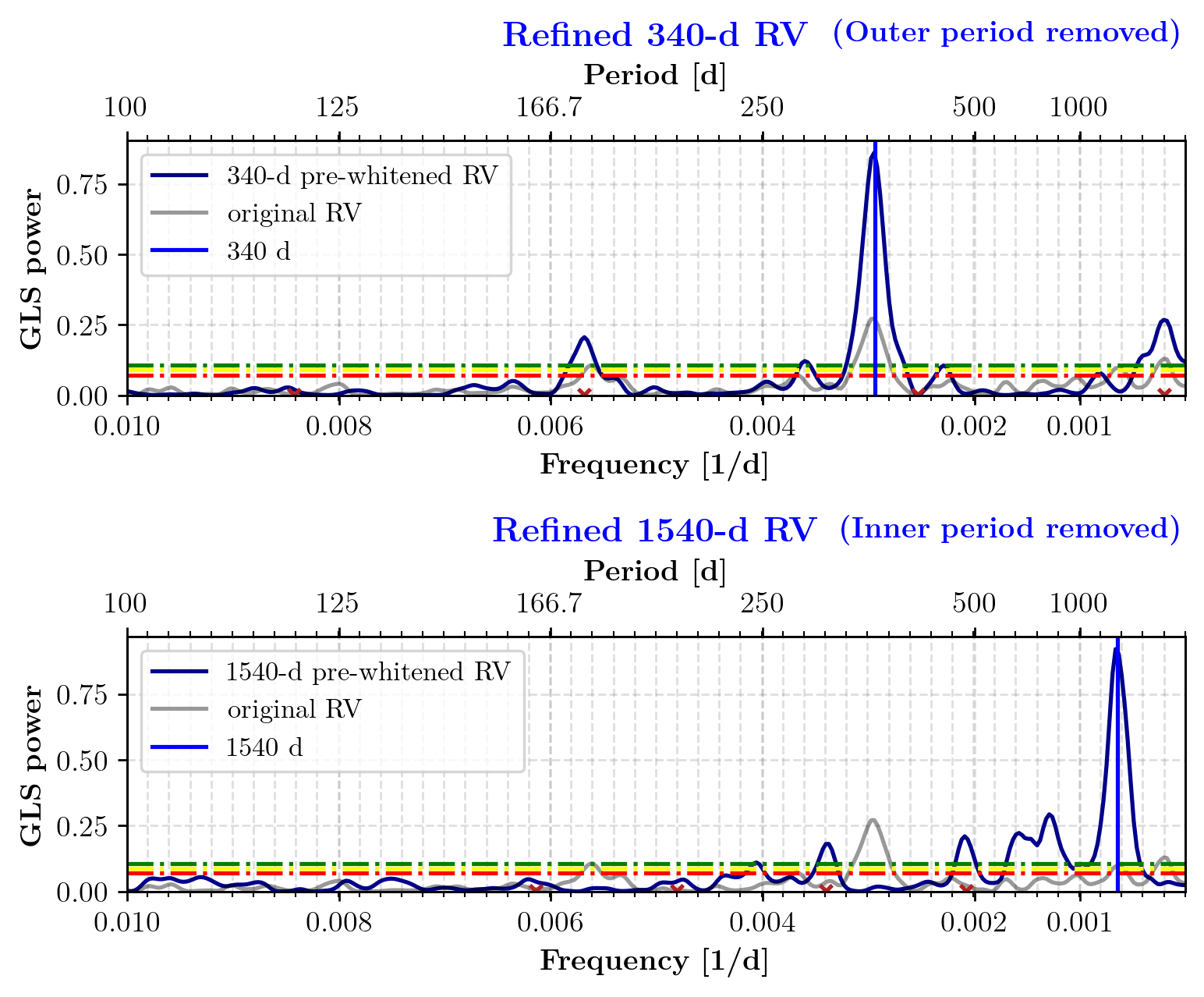}
    \caption{GLS-Ps of {pre-whitened} RV signals {in frequency space} of 2~Dra (darkblue) after {pre-whitening method} in 2-planet model, compared with the original RV (grey). \textit{Top panel} is the 340-d {pre-whitened} RV signal (after 1540-d RV removed). \textit{Bottom panel} is the 1540-d {pre-whitened} RV signal (after {340-d RV} removed). The 0.1\% (green), 1\% (yellow), and 10\% FAP (red) levels are {indicated by} dot-dash lines in each {pre-whitened} indicator, respectively. The $\times$ marker is the alias of {pre-whitened} signal indicated at {zero power (by alias relation as the same as Figure \ref{fig: GLS_p_rv_2Dra})}. 
    }
    \label{fig: Refined_GLS-P_RV}
\end{figure}
We adopted a \textit{pre-whitening} process \citep{vanicek1971} to extract the most significant and following significant peaks in GLS-P, jointly with the correlation analysis. In each iteration, we located the strongest GLS-P peak, fitted and subtracted (\textit{pre-whitened}) a Kepler model from the RVs; we recomputed the GLS-P on residuals until no significant peaks remained. Pre-whitened RV signals, as an example, were shown in Figure \ref{fig: Refined_GLS-P_RV}. We then used the resulting pre-whitened signals to compute correlation matrices between RV and indicators of stellar activity, instrument systematics, and sampling, thereby attributing each significant pre-whitened signal to its most likely origin, particularly around one-year periods. These correlation results were presented in Sections \ref{subsubsec: stellar_corr_analysis} and \ref{subsubsec: IP_corr_analysis}.  

Figure \ref{fig: GLS_p_rv_2Dra} shows the GLS-Ps of RV variations of 2~Dra with RV residuals and WF from all instruments, HIDES-S, -F1, and -F2, respectively. From 257 RV datapoints, we detected the significant 340-d RV period with FAP well below $0.1\%$ in both HIDES-S and -F1. Conversely, a WF has its dominant power at 1yr with clear separation from 340-d in all instruments, reducing the likelihood of a pure 340-d alias. However, no HIDES-F2 variations are deemed significant in any indicators due to the limited observation timescale $\lesssim 2 \text{ yr}$ and the earthquake during 2019, causing CCD order-shifts in HIDES-F2 \citep{teng2023revisiting}. We explicitly tested the HIDES-F2 sampling by fitting our best Kepler orbit at the HIDES-F2 timespan with adding errors, jitters, and computing each GLS-P; no 340-d peak was recovered in these simulations. The lack of a clear 340-d RV peak in HIDES-F2 is thus likely caused by a wide 1yr sampling from the short baseline and limited phase coverage, effectively hiding the signal. For the outer period, a 1540-d RV signal weakly appears (between $\sim0.1 \text{ - } 1\% \,\text{ FAP}$) only in the HIDES-F1's RV and RV residual when removing the 340-d signal (i.e., in $\rm{O-C}_{1}$). This outer signal is also apparent to be a long-period harmonic of $\sim5000$ days (at $\times$ marker by an alias relation \citep{dumusque2012earth}). Besides, WF signals arise in significant 1-yr and 1520-d periods in all HIDES-S, -F1, and -F2, indicating the long-term alias/harmonic influence on the outer period rather than the genuine signal.

Figure \ref{fig: SBGLS_rv} exhibits the SBGLS-P of 2~Dra RV variability, revealing the significance of RV signal evolution with highly significant probability and growing $S/N$ ratio over time ($\sim10$). The 340-d RV period is distinctive and coherently significant over observation time with increasing detection probability ($\log(\text{power})\simeq 500$). Conversely, a 1540-d RV period is far weaker than the main period and is modulated around $\log(\text{power})\sim 200$, suggesting a weak outer signal. This can be seen similarly in Figure \ref{fig: Refined_GLS-P_RV} displaying 2~Dra's RV signals after the pre-whitening process (in frequency space). A 340-d RV pre-whitened signal is sharply significant ($\ll 0.1\% \text{ FAP}$), along with its aliases in symmetric shape at $\lesssim0.5$ yr and $\sim5000$-d periods (at $\times$ markers). When pre-whitening an outer period, a 1540-d pre-whitened signal also induces a couple of several hundred-day aliases with strong significances ($<0.1\% \text{ FAP}$), assuring the long-period harmonics over $\sim$1000-d periods for 2~Dra variation.

\section{Determining the nature of the signal}
\label{sec: determine_nature_signal}
\subsection{Stellar Activity Analysis} 
\label{subsec: stellar_activity}
\subsubsection{Line profile analysis \& Chromospheric activity}
\label{subsubsec: line_profile_chrone-sph_activity}
\begin{figure}[ht!]
    \centering
    \includegraphics[width=1.\linewidth]{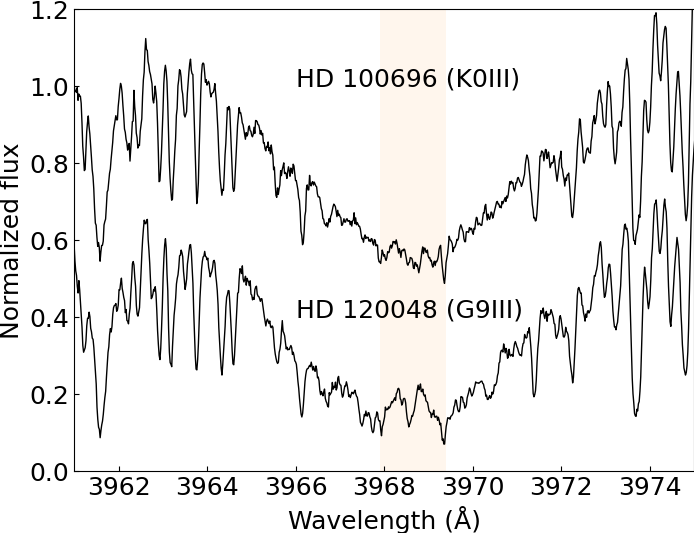}
    \caption{Ca \textsc{ii} H line spectra of quiet 2~Dra (\textit{top}) and active late-G-giant (HD120048; \textit{bottom}) during the chromospheric emission at core lines (yellow band).
    }
    \label{fig:  CaII_H_line}
\end{figure}
\begin{figure*}[ht!]
    \centering
    \includegraphics[width=0.95\linewidth]{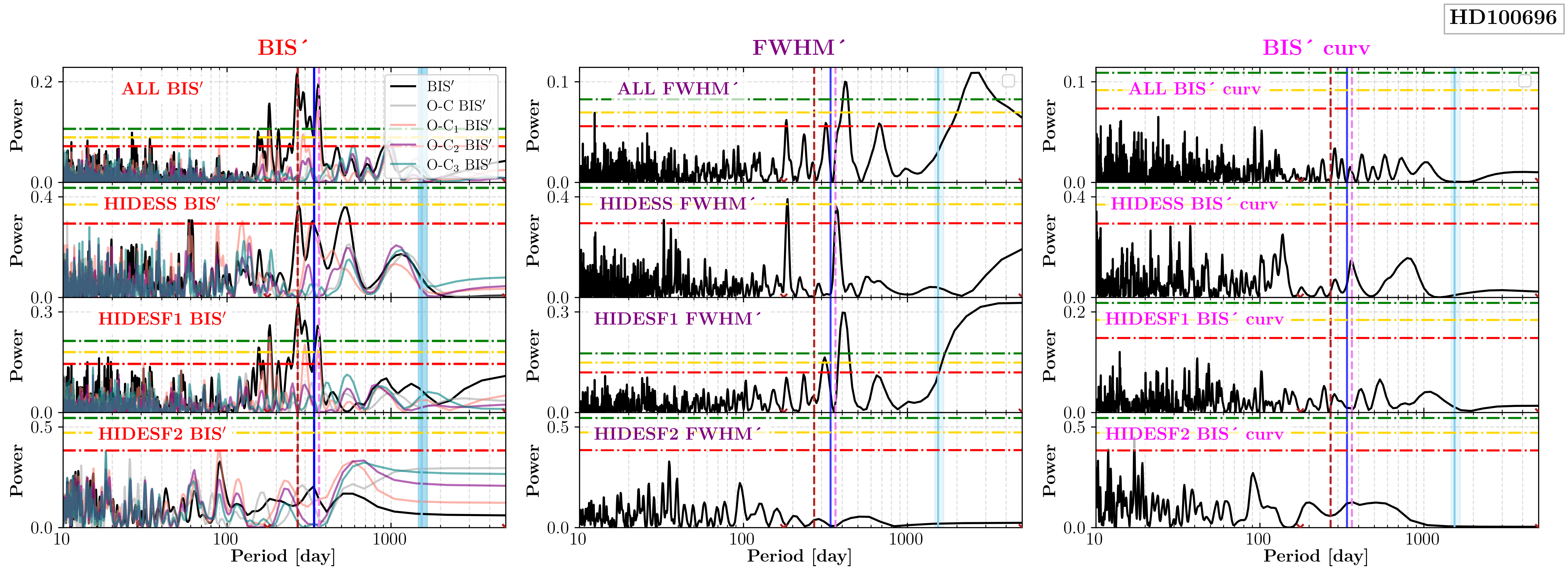}
    \caption{GLS-Ps of stellar variations of 2~Dra: BIS$'$ (\textit{left panels}), FWHM$'$ (\textit{middle panels}), and BIS$'$ curv (\textit{right panels}) (while $S_{H}$ in Figure \ref{fig: GLS_p_rv_2Dra}). Ordered BIS$'$ residuals follow the removing BIS$'$ peaks in the significant order (e.g., $\rm{O-C}_{1}$ = 270-d removed (\textit{orange}), $\rm{O-C}_{2}$ = 360-d removed (\textit{violet}), $\rm{O-C}_{3}$ = 290-d removed (\textit{teal}), and $\rm{O-C}$ = 320-d removed (\textit{grey}); \textit{see text}). Significant peak lines (e.g, vertical blue (340 d), skyblue (1540 d), brown dashed (270 d), and pink dashed lines (1 yr)), FAP levels, and alias markers are in similar details to those in Figure \ref{fig: GLS_p_rv_2Dra}.
    }
    \label{fig: GLS_p_stellar_2Dra}
\end{figure*}
\begin{figure}[!ht]
    \centering
    \includegraphics[width=1\linewidth]{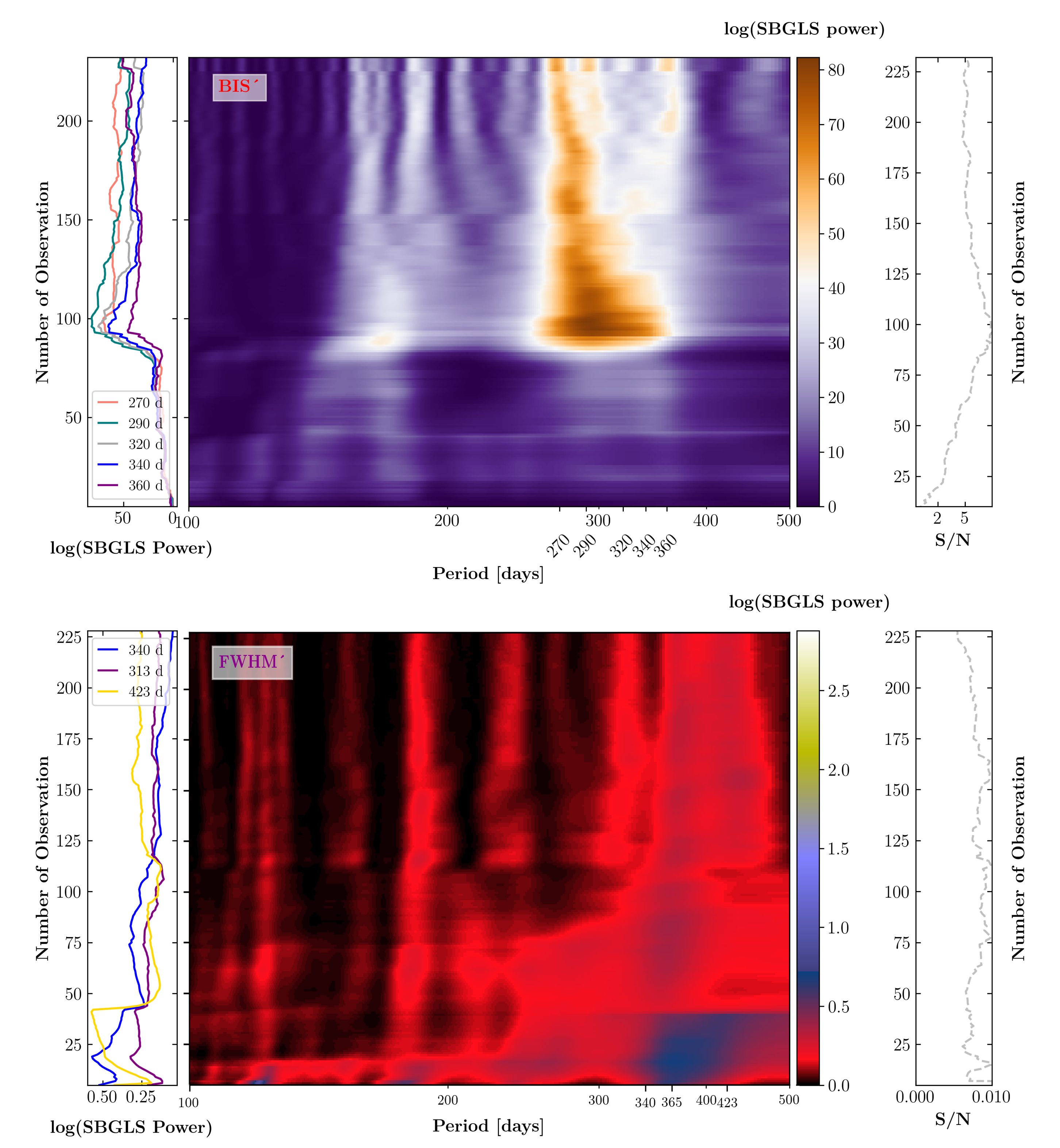}
    \caption{SBGLS-Ps of stellar variations of 2~Dra: $\text{BIS}'$ (\textit{top panels}), and $\text{FWHM}'$ (\textit{bottom panels}). Left, central, and right subplots provide identical details to SBGLS-P of RV in Figure \ref{fig: SBGLS_rv}.
    }
    \label{fig: SBGLS_bis_fwhm}
\end{figure}
With stellar variation, the deformation of the line profile can shift the spectrum's wavelength, producing the line profile asymmetry and RV-mimicked signal. We used bisector inverse span (BIS; \cite{queloz2001no}) and BIS curvature (BIS curv; \cite{sato2012double}) as stellar indicators to trace the asymmetry in cross correlation function (CCF) profile \citep{takarada2018planets,teng2022regular}. Our line analysis employed the $I_{2}$-free stellar spectra in 4000-5000 \AA \ and the numerical mask of the G-type giant model \texttt{SPECTRUM} \citep{gray1994calibration} to compute the weighted CCF \citep{pepe2002coralie}. The CCF profile exhibits the mean absorption line at each central wavelength (velocity) as the resulting RV value. We derived the BIS and BIS curvature with errors in different velocity spans of the CCF profile. The BIS (in \cite{dall2006bisectors}) is defined as
\begin{equation}
    \text{BIS} = v_{\text{top}} - v_{\text{bottom}},
    \label{eq: bis}
\end{equation}
where $v_{\text{top}}$ and $v_{\text{bottom}}$ are the mean velocities of $5-15\%$ (upper region) and $85-95\%$ (lower region) of the absorption line depth of CCF bisectors, respectively. BIS curvature represents the stellar convection and granulation by measuring the non-linearity of the bisector shape (C-shape) distortion, as the $2^{\text{nd}}$-order velocity difference between the upper and lower bisectors, i.e.
\begin{align}
    &\text{BIS curv} = (v_{\text{top}} - v_{\text{mid-top}}) - (v_{\text{mid-top}} - v_{\text{bottom}}),
    \label{eq: bis_curv}
\end{align}
where $v_{\text{mid-top}}$ is the mean velocity of $25-35\%$ of the CCF profile. Similarly, the full-width-half-maximum (FWHM) of CCF bisectors characterizes the rotation rate of the host star, which served as another stellar indicator for the stellar line broadening. Because the line broadening of slow-rotating giants is much slower than the instrumental broadening, inducing the instrumental offset, we removed the instrumental offsets in BIS, BIS curvature, and FWHM, by subtracting their respective mean values to evaluate their intrinsic stellar variations, defined as:  
\begin{subequations}
  \begin{align}
    \text{BIS}' &= \text{BIS} - \overline{\text{BIS}}, 
    \label{eq: bis_pri}\\
    \text{BIS}'\text{curv} &= \text{BIS curv} - \overline{\text{BIS curv}}, \label{eq: bis_pri_curv}\\
    \text{FWHM}' &= \text{FWHM} - \overline{\text{FWHM}}, \label{eq: fwhm_pri}
  \end{align}
\end{subequations}
where the overline denotes the mean (offset) values of each instrument, and outliers were screened out by the interquartile range ($\text{IQR} = Q_{3} - Q_{1}$) method of weighted quantile with Tukey’s 1.5-IQR outlier threshold: $[Q_{1} - 1.5 \cdot \text{IQR}, \ Q_{3} + 1.5 \cdot \text{IQR}]$. 
Chromospheric emission can also manifest line deformation and masquerade as an RV signal. To check whether our RV variations are activity-driven by the stellar chromosphere, we examined their correlation with the Ca \textsc{ii} H\&K chromospheric index (at 3968.47 \AA, 3933 \AA ). Since the 3700-4000 \AA \ spectra of HIDES-F1 and HIDES-F2 were compromised by severe aperture overlaps -- and the Ca \textsc{ii} K lines of HIDES-S spectra were too weak -- we adopted only the Ca \textsc{ii} H lines from HIDES-S spectra. According to \citet{sato2013planetary}, the Ca  \textsc{ii} H index (i.e., $S_{H}$) is defined as
\begin{equation}
    S_{H} = \dfrac{F_{H}}{F_{B} + F_{R}},
    \label{eq: sh}
\end{equation}
where $F_{H}$ denotes the integrated Ca \textsc{ii} flux at the centered H line ($\lambda_{S_{H}} = 3968.47 \ $\AA ) within a 0.66 \AA \ wavelength range, and $F_{B}, F_{R}$ are the integrated Ca \textsc{ii} fluxes at the lower and higher wavelengths from the centered H line ($\lambda_{S_{H}} -1.2 \ $\AA \ and $\lambda_{S_{H}} + 1.2 \ $\AA) within a 1.1 \AA \ wavelength range, respectively. The $S_{H}$ errors were determined from the photon noise at the H lines, and their outliers were removed in the same way as other indicators. Following in Figure \ref{fig:  CaII_H_line}, no chromospheric emission at the Ca \textsc{ii} H lines is significant in 2~Dra, representing the chromospherically quiet star.

With clean offsets and outliers, $\text{BIS}', \text{BIS}'\text{curv},  \text{FWHM}'$, and $S_{H}$ {served as} stellar indicators for line deformation and broadening. {These were shown} in { the GLS-Ps and} SBGLS-Ps (in Figures \ref{fig: GLS_p_stellar_2Dra} and \ref{fig: SBGLS_bis_fwhm}, Section \ref{subsubsec: activity_periodogram_analysis}) and in the correlation maps for both pre-whitened RV signals (Figure \ref{fig: sum_indicator_corr_matrix}) and RV data (Figure \ref{fig: bis_corr}, Section \ref{subsubsec: stellar_corr_analysis}).

\subsubsection{Activity periodogram analysis}
\label{subsubsec: activity_periodogram_analysis}
Figure \ref{fig: GLS_p_stellar_2Dra} shows the GLS-Ps of stellar variations of 2~Dra by BIS$'$, FWHM$'$, and BIS$'$ curv, respectively. BIS$'$ variations exhibit in 270-d, 290-d, 320-d, and 360-d periods. Similarly, FWHM$'$ variations are present in 313-d and 423-d periods, while BIS$'$ curv has no significant variation. Each significant BIS$'$ peak was removed via a pre-whitening method from BIS residuals following a significant order. For major signals, $\text{O - C}_{1, \text{ BIS}'}$ (\textit{orange} line) and $\text{O - C}_{2, \text{ BIS}'}$ (\textit{violet} line) show the removed 270-d and 360-d BIS$'$ peaks. For minor peaks, $\text{O - C}_{3,  \text{ BIS}'}$ (\textit{teal} line) and $\text{O - C}_{ \text{ BIS}'} $ (\textit{grey} line) present the removed 290-d and 320-d BIS$'$ peaks, respectively. Both 290-d and 320-d $\text{BIS}'$ signals may be indicated as aliases, which were weakened after subtracting 1-yr sinusoidal fit under $\sim$0.1\% and 10\% FAP levels, respectively \citep{dumusque2015characterization}. 360-d $\text{BIS}'$ signal is also likely a yearly alias, due to strong correlation with 1-yr WF signal, indicating the sampling effects in many BIS$'$ periods. However, the 270-d $\text{BIS}'$ signal does not essentially correlate with 1-yr WF and has little effect when subtracting a 1-yr sinusoidal fit, suggesting the stellar source rather than a sampling/harmonic.

We assessed stellar variations near 340-d periods in the HIDES-S (e.g., BIS$'$ and FWHM$'$). The 360-d FWHM$'$ HIDES-S variation is characterized by a yearly sampling via subtracting a 1-yr sinusoidal fit with a strong 1-yr WF correlation. However, $\sim$340-d BIS$'$ signal in HIDES-S is likely produced by the superposition of the 290-d/320-d BIS$'$ signals and the yearly alias. To test this, we removed the 360-d HIDES-S BIS$'$ signal (in $\text{O - C}_{1, \text{ BIS}'}$) and computed its correlation with WF; the removed 360-d BIS$'$ signal correlates strongly with the 1-yr WF signal.
After removed, the remaining BIS$'$ residuals (after $\text{O - C}_{2, \text{ BIS}'}$) show no further $\sim$340-d power in HIDES-S (or similarly in HIDES-F1). We may thus infer the $\sim$340-d BIS$'$ signal in HIDES-S as the mixed aliases around 1-yr, instead of a sole stellar origin.

Figure \ref{fig: SBGLS_bis_fwhm} shows SBGLS-Ps of stellar variations of 2~Dra, with a cluster of significant BIS$'$ periods at 270-360 d ($\log(\text{power})\simeq {50}$) overlapping with the 340-d RV signal. However, $\text{FWHM}'$ exhibits weak periods during 313-423 d ($\log(\text{power})\lesssim {0.5}$), suggesting different physical origins. BIS$'$ heatmap displays the most concentrated power (\textit{darkbrown} region) in the 270-360 day periods during 80-100 observations with the strongest $S/N$ ratios ($\sim5$). In contrast, FWHM$'$ heatmap reveals the weak significance of rotation signals dissipating over time (in both power and $S/N$), particularly at the 340-d period, inferring an unlikely 340-d long-period rotation. This interpretation is further supported by no spot signatures in \textit{TESS} photometry and no strong chromospheric activity in $S_{H}$. However, since observing in short-period sector
and combining multi-sector lightcurves can be limited by sector-to-sector systematics (e.g., scattered light), a low-amplitude 340-d photometric variation would be difficult to confirm even if present.
A spot-induced RV amplitude of around $15\,\mathrm{m\,s^{-1}}$, corresponding to a spot size of $\sim1.3\%$
\footnote{$K_{\text{spot}} = (8.6\ v\sin i - 1.6) (A_{\text{spot}}/A_{\text{star}})^{0.9}$ \citep{Hatzes2002starspots}, where $K_{\text{spot}}$ is the spot-induced RV amplitude (in $\mathrm{m \, s}^{-1}$), $A_{\text{spot}}/A_{\text{star}}$ is the spot area coverage (in \%) of the star, and $v\sin i$ is in Section \ref{sec: stellar_model}.}, would typically {only} {induce flux modulations of a few $\sim0.1\%$. Though such modulations are principally observable with ground-based telescopes, \textit{TESS} light curves reveal only solar-like $p$-mode oscillations ($\nu_{\text{max}} \simeq 36\,\mu\text{Hz}$, corresponding to 0.32\,d) and show no detectable spot modulation. Both findings} {of \textit{TESS} and quiet $S_{H}$ activity} are thus against rotational-modulated active regions or spot-driven RV variations as the 340-d origin. Additionally, no long-period photometric modulations are present in either the Hipparcos or ASAS-SN light curves. Neither the day-timescale $p$-modes nor large persistent spot modulation can adequately account for the observed multi-100-day BIS$'$ variations.

However, these stellar evolutions suggest insights for activity influence over time. Even in an inactive star, the BIS$'$ clustered periods (270-360 d, Figure \ref{fig: SBGLS_bis_fwhm}) may contaminate the 340-d RV signal at specific timescales ($N_{\rm obs}\sim100$). This may also show why in Figure \ref{fig: SBGLS_rv} the 340-d RV signal is weakened and modulated during the $\sim100$ observations, then becomes strong later, when 270-360-d BIS$'$ signals are less significant over time ($N_{\rm obs}>100$). The episodic nature of the BIS$'$ signals contrasts with the persistent 340-d RV signal throughout the observation baseline. This indicates that while stellar activity may contribute to variations at specific epochs, it may not fully account for the coherent 340-d period. We thus assessed the periodic tests of stellar signals further between 270-d BIS$'$ and 340-d RV periods through phase stability analysis (Section \ref{subsubsec: phase_stability_analysis}) to determine if both signals are of the same origin.

\subsubsection{Stellar correlation analysis}
\label{subsubsec: stellar_corr_analysis}
\begin{figure*}[ht!]
    \centering
    \includegraphics[width=1\linewidth]{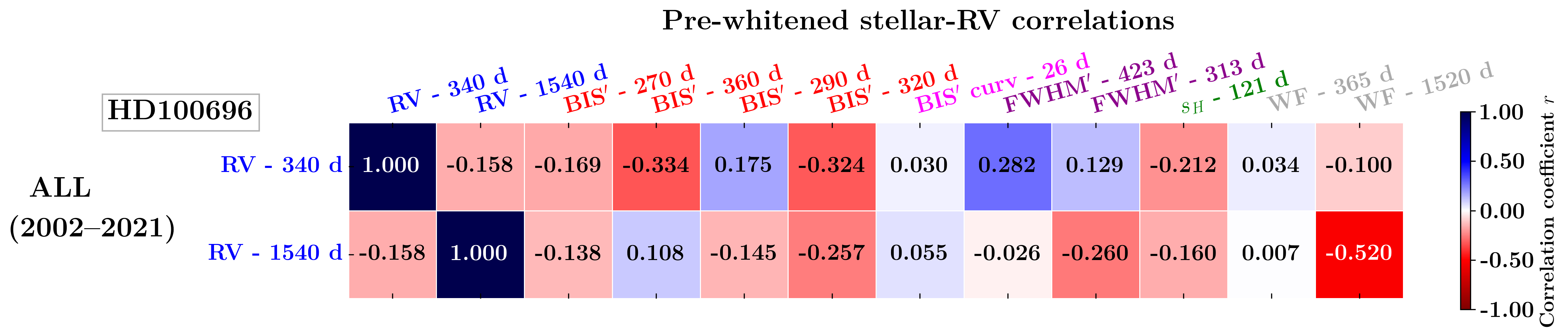}
    \caption{
    Period-confined (pre-whitened) correlations of 2~Dra's RV signals between stellar variations (e.g., $\text{BIS}', \text{BIS}'\text{curv}, \text{FWHM}', S_{H}$), and yearly variations (WF, resp.) with their significant periods (following significant peak order) in all combined instruments. 
    }
    \label{fig: sum_indicator_corr_matrix}
\end{figure*} 
\begin{figure}[ht!]
    \centering
    \includegraphics[width=1.\linewidth]{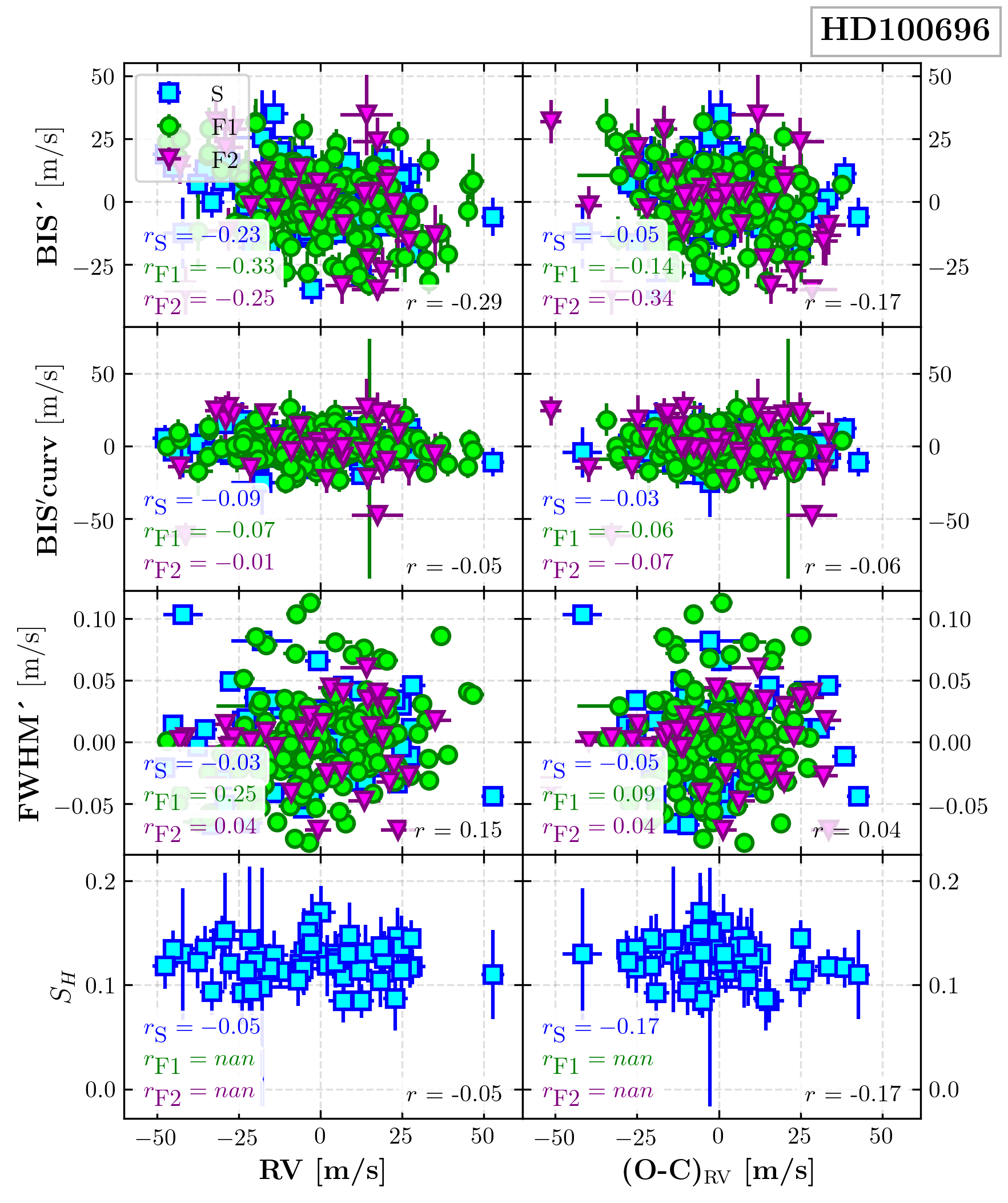}
    \caption{Stellar indicator correlation map of 2~Dra with RVs (\textit{left column}) and RV residuals (\textit{right column}). Unconfined-period correlation $r$ is present in each indicator pair in the bottom right of each panel, while comparing $r$ of the indicator pair in each HIDES (e.g., -S, -F1, and -F2) is indicated in the bottom left of each panel, respectively.
    }
    \label{fig: bis_corr}
\end{figure}
Since the significant periods in the RV (340 d) and the activity indicators (270-360 d BIS$'$) are very close to each other and to the 1yr alias, we performed the correlation analysis on the pre-whitened signals (Figure \ref{fig: sum_indicator_corr_matrix}) and RV-activity data (Figure \ref{fig: bis_corr}). These help to determine if these signals are intrinsically correlated or influence each other or not.
Even {their significant} periods are constrained {(or pre-whitened, in 121-423 d in Figure \ref{fig: sum_indicator_corr_matrix})}. {All pre-whitened} $\text{BIS}', \text{BIS}'\text{curv}, \text{FWHM}'$, and $S_{H}$ {signals} {still weakly correlate} to the 340-d RV signal {($|r| \simeq 0.03\text{ - }0.33$)}. {Comparatively,} Figure \ref{fig: bis_corr} shows correlation maps of stellar indicators of 2~Dra with RV and RV residuals, when no {significant} periods are constrained. All $\text{BIS}', \text{BIS}'\text{curv}, \text{FWHM}'$, and $S_{H}$ {data also} weakly correlate with both RV and RV residuals ($|r| \lesssim 0.29$). 

These stellar correlation results reveal insights for weak rotational effects. Even if {we assumed stellar rotation manifests in 270-360-d} $\text{BIS}'$ or {313-423-d} $\text{FWHM}'$ (via differential rotation or {magnetic activity}), the rotational variations are {unlikely to drive the 340-d RV signal}. This conclusion is mainly supported by the weak correlations between BIS$'$ and RV. Specifically, their correlations are weak both when analyzing the period-confined 340-d RV signal ($|r|\leq 0.33$; Figure \ref{fig: sum_indicator_corr_matrix}) and when using the non-period-confined data ($|r|\lesssim 0.29$); Figure \ref{fig: bis_corr}). Notably, $\text{BIS}{'}$-RV correlations in Figure \ref{fig: bis_corr} appear to vary slightly between instruments or timespans. Their correlation is modulatedly weak during the HIDES-S and -F2 epochs  ($r_{\text{S}} = -0.23$, $r_{\text{F2}} = -0.25$) but marginally higher during the HIDES-F1 measurements ($r_{\text{F1}} = -0.33$). This might suggest the stellar modulation evolving over time. However, the activity correlations remain weak collectively across all line indicators ($|r|\lesssim 0.29$), including FWHM$'$. The correlation analysis using pre-whitened signals (Figure \ref{fig: sum_indicator_corr_matrix}) also reinforces these findings. The 340-d RV period remains weakly correlated to all activity signals (BIS$'$, BIS$'$ curv, FWHM$'$, $S_{H}$ with $|r|\leq 0.33$) without 1yr WF correlation. Conversely, the 1540-d RV signal highly correlates to only a long-period WF ($|r| = 0.52$, thereby confirming an alias), with negligible stellar correlations. Given weak activity, the rotational variabilities are thus probably not essentially associated with the 340-d RV variation. Stellar variations to induce RV of 2~Dra may not be the dominant source via line deformation, broadening, or chromospheric emission. 

With the activity timescales $\lesssim$ a few hundreds of days (e.g., radial oscillations (${\simeq0.32 \text{ d}}$), and granulations (${\sim0.7-70 \text{ d}}$)
\footnote{
$\tau_{\text{granule}} = \tau_{\text{granule},\odot} \cdot \left(\frac{\nu_{\text{max},\odot}}{\nu_{\text{max}}}\right)$ \citep{huber2009automated}, given $\nu_{\text{max},\odot} = 3090\ \mu\text{Hz}$, $\tau_{\text{granule},\odot} = 0.2, 3, 20 \text{ hr}$ for granulation, mesogranulation, and supergranulation, respectively \citep{rast2003scales}}), no stellar variations from the rotation, $p$-mode oscillation, granulation, and spots are significantly correlated with the RV variations for 2~Dra.

\subsection{Instrumental variability} 
\label{subsec: inst_variability}
\subsubsection{Instrument profile analysis} \label{subsubsec: inst_profile_analysis}
\begin{figure}[ht!]
    \centering
    \includegraphics[width=0.95\linewidth]{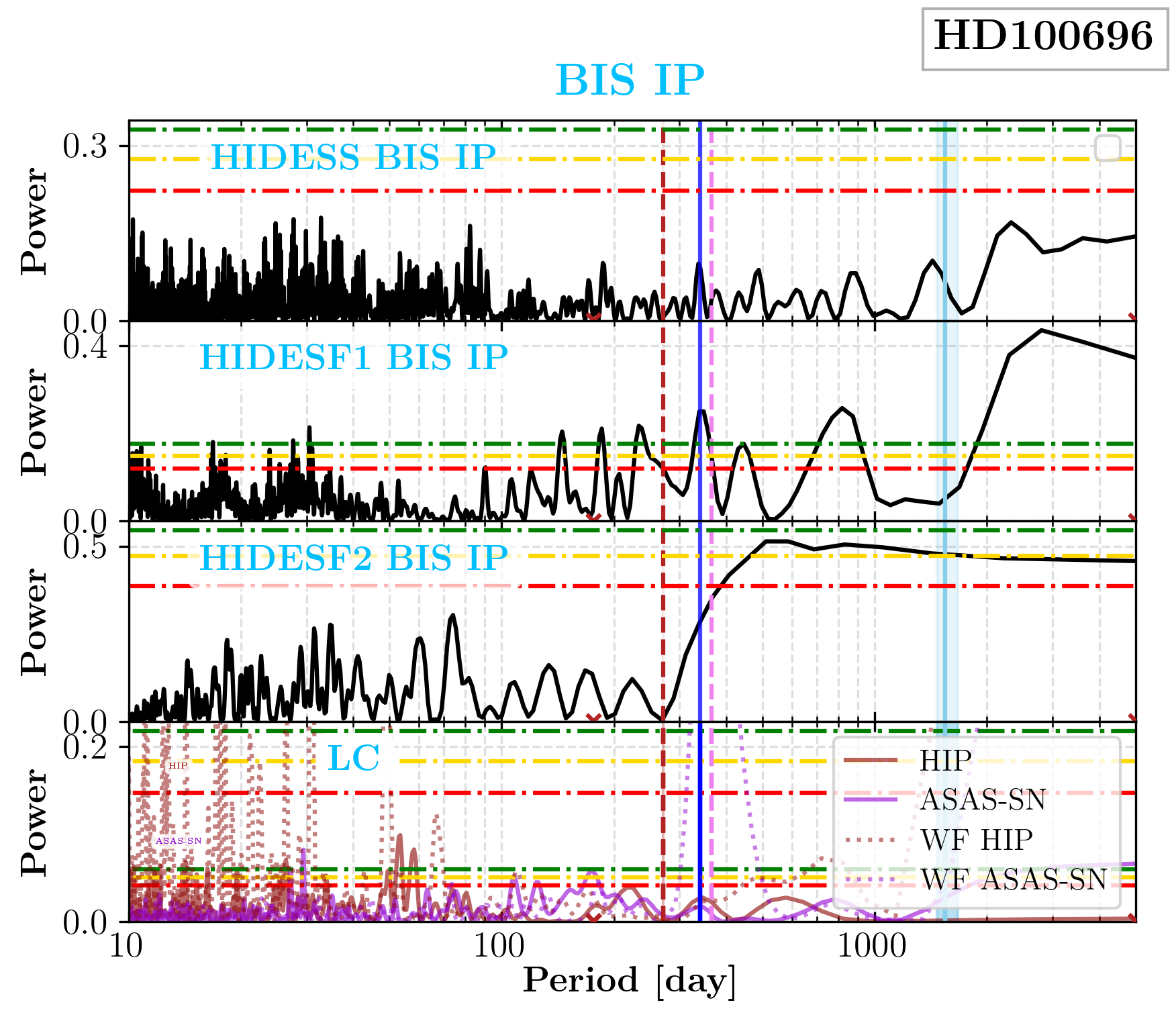}
    \caption{GLS-Ps of instrumental variations in 2~Dra: BIS IP of HIDES-S (\textit{top panel}), -F1 (\textit{second panel}), and -F2 (\textit{third panel}); and GLS-Ps of 2~Dra's lightcurve (LC) variations (\textit{bottom panel}) from Hipparcos and ASAS-SN with their WF. 
    Significant peak lines (e.g, vertical blue (340 d), skyblue (1540 d), brown dashed (270 d), and pink dashed lines (1 yr)), FAP levels, and alias markers are in similar details to those in Figure \ref{fig: GLS_p_rv_2Dra}.
    Hipparcos FAP levels are indicated at power$\sim 0.2$, and ASAS-SN FAP levels are located at power $< 0.1$.
    }
    \label{fig: GLS_p_IP_2Dra}
\end{figure}
\begin{figure}[ht!]
    \centering
    \includegraphics[width=1.\linewidth]{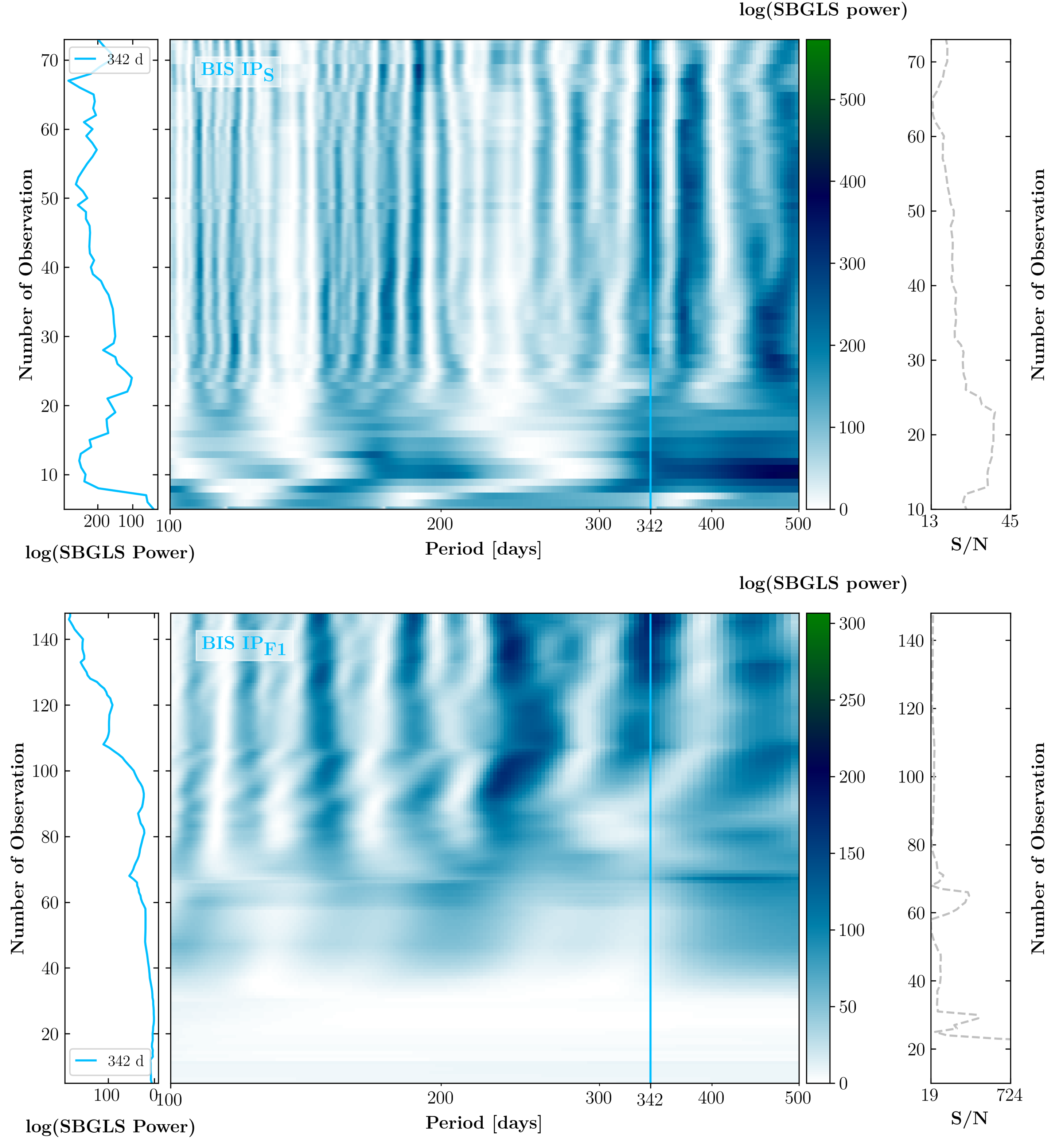}
    \caption{SBGLS-Ps of instrumental variations in 2~Dra between BIS IP of HIDES-S (\textit{top panels}) and HIDES-F1 (\textit{bottom panels}) (excl. HIDES-F2 due to a short observation baseline). Left, central, and right subplots provide identical details as Figure \ref{fig: SBGLS_rv}.
    }
    \label{fig: SBGLS_bip}
\end{figure}
Apart from stellar variation, the instrument profile (IP) can shift {spectral} line deformation through the instrumental broadening, {causing a Doppler shift by instrument.} With HIDES's velocity resolution $c/R \simeq 4.4 \text{ - } 5.4 \, \mathrm{km}\,\mathrm{s}^{-1}$ (given speed of light $c$ and resolutions $R$) higher than the rotation velocity of 2~Dra ($v\sin i = 1.68 \, \mathrm{km}\,\mathrm{s}^{-1}$), the IP variability {might have a possibility to affect an RV in each instrument.} {To track and remove RV errors caused by IP variations, we modeled HIDES spectra with the $I_2$-cell forward approach \citep{sato2002development}. In order words, the observed stellar+$I_2$ spectrum was fitted chunk-by-chunk with a wavelength- and time-dependent IP. This framework, introduced by \citet{valenti1995determining} using $I_2$-cell to determine the spectrograph’s IP and formalized by \citet{butler1996attaining}, solved simultaneously for the Doppler shift and the IP drift.} The IP model was derived {from}  {a central Gaussian with} {ten} {satellite} Gaussian profile{s following \citet{butler1996attaining, sato2002development}} based on different HIDES properties. To assess the {IP} variability, we {derived} the BIS of the mean IP (BIS IP) following \citet{takarada2018planets} approach. {We traced} the {IP} broadening (asymmetry) in different velocity spans of the instrument spectra, assessed by a chromospherically quiet G-type giant star \citep{takarada2018planets}.
Given the IP modeling, no IP–RV correlations are expected in case the IP model effectively captures IP variation. We assessed this quantitatively for 2~Dra in Section \ref{subsubsec: IP_corr_analysis}, and RV standard stars jointly with our star catalog in Section \ref{subsubsec: standard_star}.

Figure \ref{fig: GLS_p_IP_2Dra} displays the GLS-Ps of BIS IP variations between HIDES-S, -F1, and -F2 in 2~Dra measurements, and photometric variations of 2~Dra by Hipparcos and All Sky Automated Survey for SuperNovae (ASAS-SN). In HIDES-S, no significant IP variation is detected. However, BIS IP variation in the HIDES-F1 is significantly detected at the $\sim$342-d period {($\leq0.1\%  \text{ FAP}$).} But in HIDES-F2, only long-period harmonics are shown, thereby an insignificant IP variation. No lightcurve variations are also significant for the 340-d signal in 2~Dra by Hipparcos and ASAS-SN. 

However, Figure \ref{fig: SBGLS_bip} yields BIS IP variations {in SBGLS-Ps} for {both} HIDES-S and -F1 with comparable significance at $\sim$340-d periods (${\log(\rm{power}) \simeq 200}$), to the RV with a decreasing $S/N$ ratio over time. HIDES-S IP variations modulate with a significant power around the $\sim$340-d period, and HIDES-F1 IP variations are increasingly significant over time at the $\sim$340-d period, despite sharply declining $S/N$ ratios. These IP variations {might arguably imply whether they induce} RV variations at particular $\sim$340 days, or are just the instrumental artifacts around annual periods. 
 
To test this, we {thus} examined BIS IP significance {to determine} whether IP variations intrinsically influence RV signal, particularly in the 340-d period, through IP correlation analysis (Section \ref{subsubsec: IP_corr_analysis}), RV standard stars (Section \ref{subsubsec: standard_star}), and phase stability analysis (Section \ref{subsubsec: phase_stability_analysis}).

\subsubsection{IP correlation analysis}
\label{subsubsec: IP_corr_analysis}
\begin{figure}[ht!]
    \centering
    \includegraphics[width=.75\linewidth]{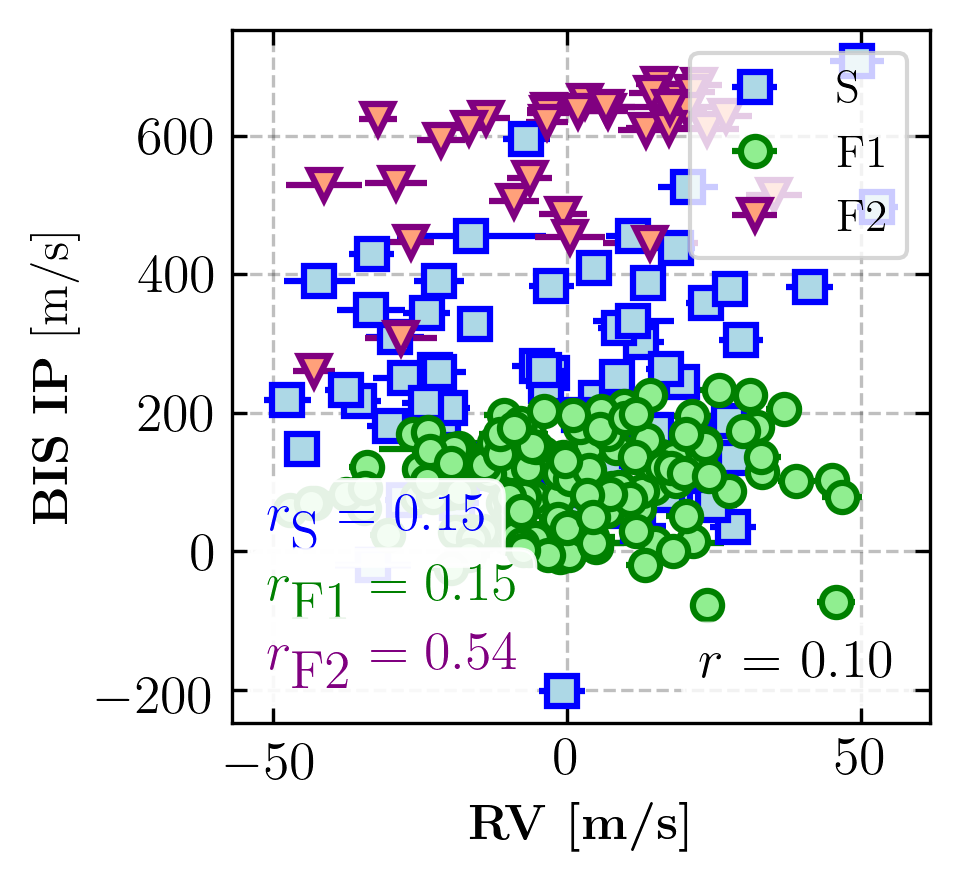}
    \caption{BIS IP correlation map of 2~Dra with RV. The correlation $r$ values provide identical details to Figure \ref{fig: bis_corr}.
    }
    \label{fig: bip_corr_map}
\end{figure}
Figure~\ref{fig: bip_corr_map} exhibits BIS IP correlation map with RV for 2~Dra, {showing} no significant correlations with the RV variations {across HIDES-S and -F1 ($|r| \lesssim 0.15$)}. {These negligible correlations are expected from the IP model for HIDES-S and -F1, indicating that RV errors associated with IP changes are effectively removed. By contrast, HIDES-F2 exhibits a substantially larger IP–RV correlation ($r \simeq 0.5$). This elevated IP correlation is plausibly attributable to post-earthquake mechanical/optical changes and a relatively short, sparsely-sampled F2 baseline. These consequently may leak IP residuals into RVs that the $I_2$-cell forward model fails to remove, and IP residuals could highly correlate with RVs. HIDES-F2 IP variation, despite a high IP-RV correlation, is unfortunately inconsequential for correlation analysis as a result of the earthquake incidents, short timescale, and long-period harmonics.} With barely IP correlations to RV from {HIDES-S and -F1} instruments, the IP drifts from HIDES-S and -F1 likely have no intrinsic impact on the 340-d RV signal, probably caused by latent annual systematics, particularly in HIDES-F1, leading to the IP leakages around 340-d periods.

\subsubsection{RV standard stars}
\label{subsubsec: standard_star}
\begin{figure*}[!ht]
    \centering
    \includegraphics[width=\linewidth]{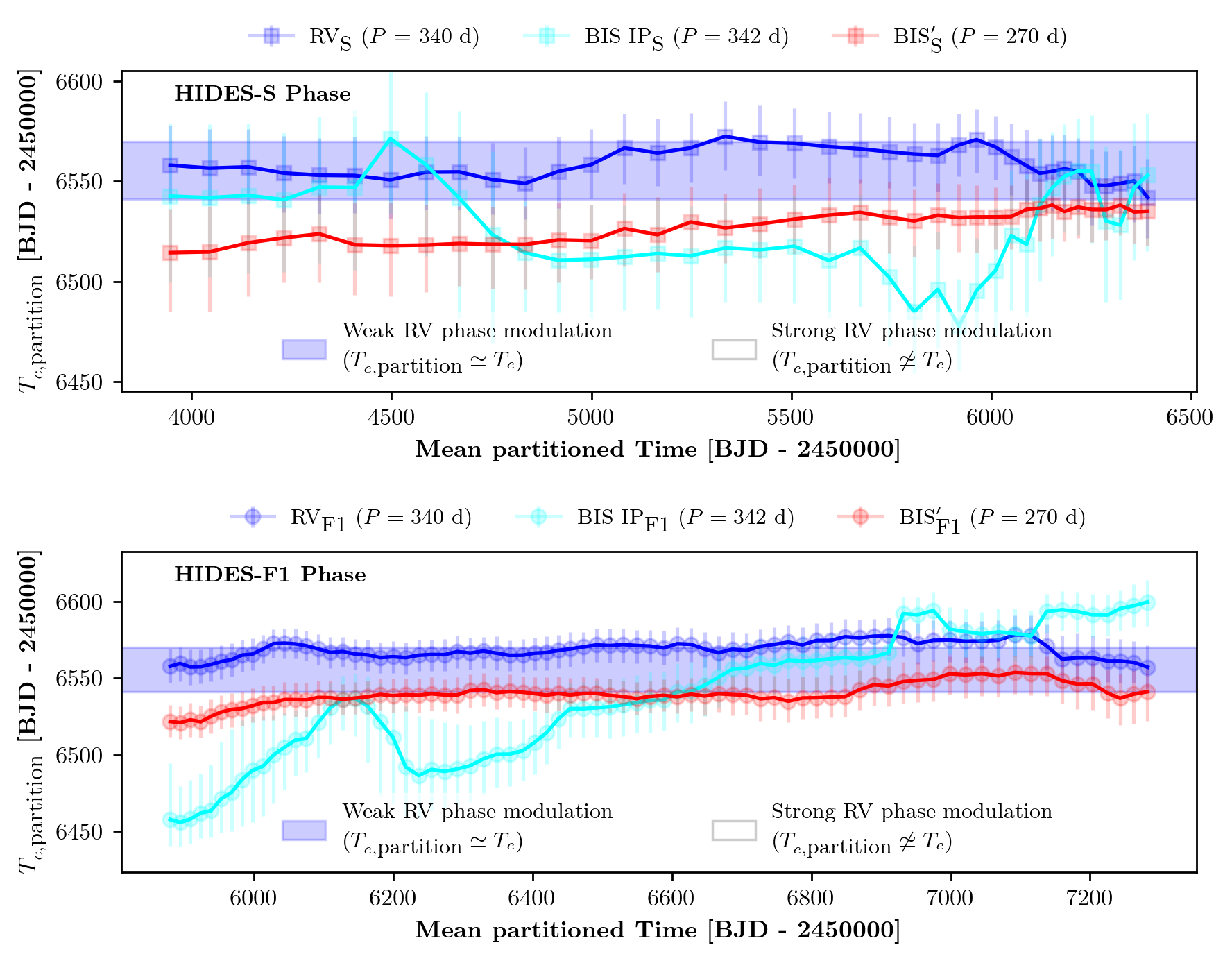}
    \caption{Phase {partitions} of 2~Dra signals. {The modulation stability} is indicated by {relative flatness} {across partitions} in the \textit{weak modulation zones} between HIDES-S (\textit{top panel}) and HIDES-F1 \textit{(bottom panel}) measurements. 
    Each panel shows the {partitioned} phase ($T_{c}$) evolution of HIDES-S and -F1's indicators in each partitioned time between 340-d RV (\textit{blue}), 342-d BIS IP (\textit{cyan}), and 270-d $\text{BIS}'$ variations \textit{(red)}, respectively.
    Blue region is the posterior $T_{c}$ with $1\sigma$ errorband of full RV data from Table \ref{tab: orbital_param} (1-planet model),
    suggesting the \textit{weak RV phase modulation zone} for the RV's partitioned $T_{c}$.
    When partitioned $T_{c}$ is fully outside of the blue region, significant RV phase modulations may be implied.
    }
    \label{fig: phase_stability_2Dra_SS_F1}
\end{figure*}
In our RV modeling, we employed two RV standard stars, $\tau$~Cet (HD~10700) and $\epsilon$~Vir (HD~113226), as reference targets to assess the IP stability of our measurements.

For the first standard star, $\tau$~Cet, no evidence of a 340-d RV signal was found in HIDES-S or{-F1, according to \citet[(in their Fig. 11)]{teng2022regular}}, although HIDES-F2 revealed yearly RV variations close to 340 d \citep{teng2023revisiting} (in their Fig. 43). Notably, no 340-d BIS IP variation appears in any HIDES instrument. For the second standard star, $\epsilon$~Vir, HIDES-S and -F1 likewise show no significant RV variability at either 340~d or 1~yr \citep{teng2022regular}, and HIDES-F2 exhibits no significant RV or BIS IP power near 340~d \citep{teng2023revisiting}.

Within our EAPS-Net\footnote{East-Asian Planet Search Network \citep{izumiura2005east}} catalog, however, we identified $\sim$340-d BIS IP variations in certain stars observed with HIDES-F1. These exhibit negligible RV correlations ($|r|\sim 0.1 \text{ - } 0.2$), and only a few show at most moderate IP–RV correlations ($|r|\sim0.3\text{ - }0.5$). We infer that these $\sim$340-d modulations likely arise from  \textit{i}) resonances or interference between instrumental and yearly variations across different instruments, and/or \textit{ii}) merged effects between yearly sampling and {IP} systematics. This might explain why 2~Dra’s 340-d RV signal superficially aligns with IP systematics near the same period, reflecting instrumental influences that can be triggered around 1-yr periodicities.

\subsubsection{Phase stability Analysis}
\label{subsubsec: phase_stability_analysis}
To verify whether the 340-d RV signal shares a common origin with stellar/instrument indicators, we tracked the phase ($T_{c}$) and amplitude ($K$) of each periodicity in overlapping time partitions while fixing known period $P$ (and all else fixed) in three instrument epochs (HIDES-S, -F1, and -F2). We performed Keplerian fits in shifted partitions ($\gtrsim$2000 BJD) and recovered $T_{c}$ or $K$ by \texttt{MCMC} in each partition (bounds: $T_{c}$ within one period; $K$ within $\mathcal{U}(\frac{K}{2}, \frac{3K}{2})$). HIDES-F2 was excluded because its $\leq$2-yr baseline and the 2019 CCD-order shift yielded unstable estimates for $P = 340 \text{ d}$. As diagnostic tests, the significant RV phase or amplitude modulations across epochs---fully outside the $1\sigma$ errorband of Keplerian-fitted $T_c$ or $K$ from full RV data---would indicate non-planetary nature. If the partitioning $T_c/K$ behaviors of 340-d RV (of -F1) and $\sim$340-d IP (of -F1) resembled each other, the IP would be a culprit. 

Figure \ref{fig: phase_stability_2Dra_SS_F1} displays the stability plots of phase partitions between variation sources of 2~Dra: RV (planet candidate), $\text{BIS}'$ (activity), and BIS IP (instrumental drift) across HIDES-S and -F1 instruments (with corresponding amplitude partitions provided in the Appendix). {The shaded blue regions indicate the $1\sigma$ uncertainty range of $T_c$ (and $K$ in Appendix) derived from fitting the full RV dataset, serving as reference zones for weak modulation.} {Phase-wise, the RV $T_{c}$ series ($P = 340 \text{ d}$) in HIDES-S and -F1 remains relatively stable and consistent throughout most time partitions, falling within or near the $1\sigma$ reference zone. This reveals a mainly coherent signal in phase in both instruments, supporting a planet candidate origin rather than IP artifacts. The BIS$'$ $T_{c}$ series ($P = 270 \text{ d}$) displays similar phase stability to the RV $T_{c}$, suggesting a stellar activity origin. In contrast, the BIS-IP $T_{c}$ series ($P \simeq 342 \text{ d}$) shows unstable and erratic phases over time, representing instrumental characteristics.} Amplitude-wise, the RV $K$ series ($P = 340 \text{ d}$) remains largely consistent in amplitude across partitions in both HIDES-S and -F1, within or near the $1\sigma$ band. The BIS$'$ $K$ series ($P = 270 \text{ d}$) also shows relatively stable activity, paralleling RV behavior. Conversely, the BIS-IP $K$ series ($P \simeq 342 \text{ d}$) exhibits high variability and inconsistency across partitions with substantial scatter over time. These phase-amplitude diagnostics support the conclusion that the 340-d RV signal is primarily coherent across intrinsic variation, and therefore not significantly driven by IP systematics.

Although the 340-d RV signal is primarily stable, it exhibits modest localized modulations in both phase and amplitude. A noticeable phase modulation occurs during approximately $\sim$6800--7100 BJD in the HIDES-F1 data, where the phase shifts slightly beyond the $1\sigma$ reference zone. As shown in the Appendix, the amplitude ($K$ series) of the 340-d RV signal also shows localized modulations. Specifically, the RV amplitudes increase in both HIDES-S ($\sim$5900--6400 BJD) and HIDES-F1 ($\sim$5900--6000 BJD), then later modestly decrease (after $\sim$6800 BJD) relatively in the $1\sigma$ band. These localized modulations in the 340-d RV signal could indicate either intrinsic stellar variability (similar to the 270-d $\text{BIS}'$ signal) or reflect limitations of the partitioning analysis due to sparse or locally clustered data coverage. Despite these localized variations, the 340-d RV signal maintains overall phase and amplitude coherence across both instruments. The predominant consistent RV phases and amplitudes, as opposed to IP behaviors, support an astrophysical origin for the 340-d signal—either a planetary candidate with potential weak activity-induced modulation, or a long-lived intrinsic stellar activity, but not a pure IP artifact.

\enlargethispage{\baselineskip}

\section{Kepler-orbit fitting}  
\label{sec: kep_orbit_fit}
\begin{figure}[!ht]
    \centering
    \includegraphics[width=0.95\linewidth]{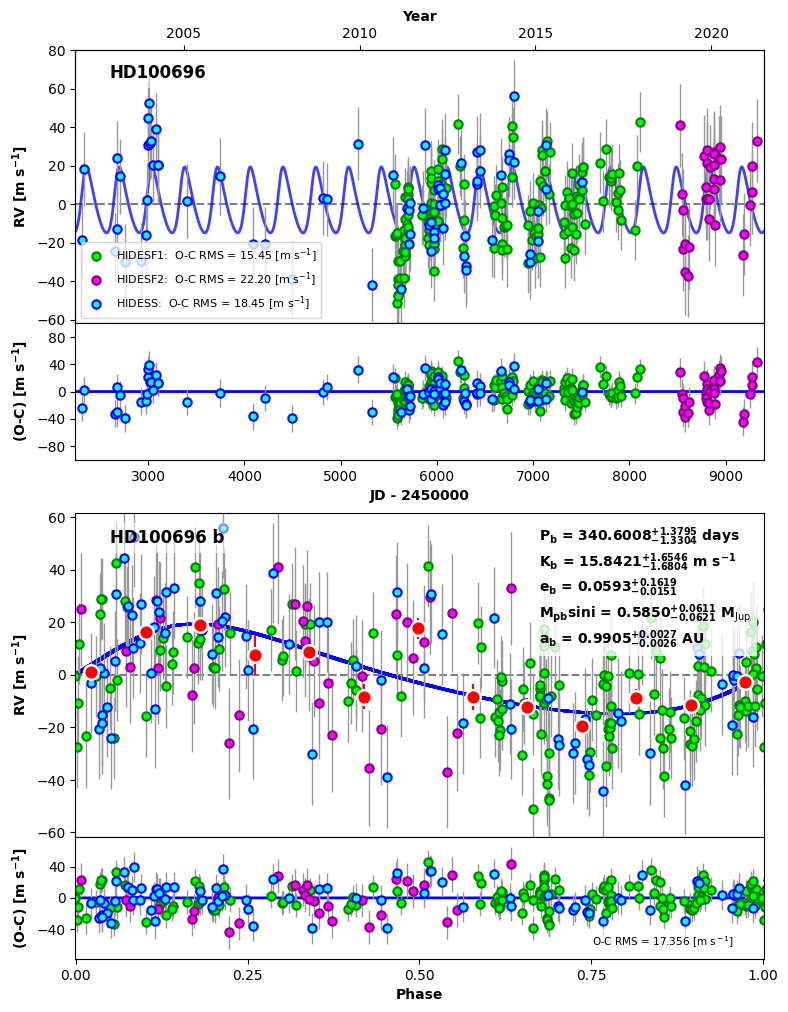}
    \caption{Hypothetical 2~Dra's 1-planet model in RV time series (\textit{top panel}) and Kepler-orbit phasefold (\textit{bottom panel}). Red bead represents each bin of RV phasefold.
    }
    \label{fig: 1PL_Kep_fit}
\end{figure}
\begin{table*}[t]
\caption{Orbital parameters of 2~Dra's planetary system}
\begin{center}
\begin{tabularx}{1.\textwidth}{lccl} 
\hline\hline
Parameters & 0-Planet & 1-Planet & Prior $\&$ Bound  \\ 
\hline\hline
Period $P$ [d] 
& - 
& $340.60_{-1.33}^{+1.38}$
& $\text{Jeffreys}(1, 3000)$ 
\\          
RV semi-amplitude $K$ [m s$^{-1}$]  
& -
& $15.84_{-1.68}^{+1.66}$ 
& $\text{Mod-Jeffreys}(1.01 (1), 1000)$
\\  
Eccentricity basis of cosine $\sqrt{e}\cos\omega$ 
& -
& $-0.048_{-0.242}^{+0.241}$ 
& $\mathcal{U}(-1, 1)$
\\          
Eccentricity basis of sine $\sqrt{e}\sin\omega$ 
& -
& $-0.239_{-0.167}^{+0.243}$ 
& $\mathcal{U}(-1, 1)$ 
\\         
Time of conjunction $T_{c}$ [BJD - 2450000] 
& -
& $6556.74_{-14.60}^{+14.55}$ 
& $\mathcal{U}(-4000, 9500)$
\\
RV offset of HIDES-S $\gamma_{\rm{S}}$ [m s$^{-1}$] 
& -
& $-2.773_{-2.260}^{+2.349}$ 
& \multirow{3}{*}{$\mathcal{U}(-800, 800)$}
\\
RV offset of HIDES-F1 $\gamma_{\rm{F1}}$ [m s$^{-1}$] & -
& $4.459_{-1.439}^{+1.470}$ 
& 
\\
RV offset of HIDES-F2 $\gamma_{\rm{F2}}$ [m s$^{-1}$] & -
& $-7.222_{-4.189}^{+4.211}$ 
& 
\\
RV jitter of HIDES-S $s_{\rm S}$ [m s$^{-1}$] 
& $22.883_{-1.823}^{+2.062}$
& $18.419_{-1.552}^{+1.730}$ 
&  \multirow{3}{*}{$\text{Mod-Jeffreys(1.01(1), 100)}$} 
\\
RV jitter of HIDES-F1 $s_{\rm F1}$ [m s$^{-1}$] 
& $18.230_{-1.038}^{+1.152}$
& $15.135_{-0.889}^{+0.988}$ 
& 
\\
RV jitter of HIDES-F2 $s_{\rm F2}$ [m s$^{-1}$] 
& $20.022_{-2.393}^{+2.878}$
& $21.830_{-2.702}^{+3.220}$ 
& 
\\
\hline
Eccentricity $e$             
& -
& $0.059_{-0.015}^{+0.162}$ 
& Derived 
\\           
Longitude of periastron $\omega$ [rad] 
& - 
& $-1.771_{-0.647}^{+1.838}$
& Derived 
\\      
Time of periastron $T_{p}$ [BJD - 2450000] 
& -
& $6713.85_{-14.60}^{+14.55}$  
& Derived 
\\
Minimum planet mass $M_{P} \sin{i}$ [$M_{\rm Jup}$] 
& -
& $0.585_{-0.062}^{+0.061}$ 
& Derived 
\\
Orbital distance $a$ [AU] 
& -
& $0.9905_{-0.0026}^{+0.0027}$  
& Derived 
\\  
rms [m s$^{-1}$] 
& 20.210
& 17.356 
& Derived 
\\ 
$\chi^{2}_{\rm red}$ 
& 0.94
& 1.01 
& Derived  \\
BIC 
& 2286.79
& 2246.93 
& Derived \\ 
\hline
\end{tabularx}
{\footnotesize \raggedright 
Note: $\text{Jeffreys}(x_{\text{min}}, x_{\text{max}}) = \ln [ (x \ln ( x_{\rm max}/x_{\rm min} ) )^{-1}  ] $.  $\text{Mod-Jeffreys}(x_{\text{min}}(a), x_{\text{max}}) = \ln [ ( (x-a) \ln (  \frac{ x_{\rm max} -x_{\rm min}}{x_{\rm min} - a} )  )^{-1}]  $.  $\mathcal{U}(x_{\text{min}}, x_{\text{max}}) = [x_{\text{max}}-x_{\text{min}}]^{-1}$\par}
\end{center}
\label{tab: orbital_param}
\end{table*}
We performed Keplerian orbit fitting for our RV model using Bayesian inference with the \texttt{radvel} package \citep{fulton2018radvel}. Our approach involved deriving maximum a posteriori (MAP) estimates through the \citet{powell2006fast}'s optimization method and sampling posterior distributions using \texttt{MCMC} for the orbital parameters. For our planet model, we adopted the following fitting parameters {as a standard basis \citep{fulton2018radvel}}: period $P$, time of transit $T_{c}$, RV semi-amplitude $K$, and eccentricity basis of longitude of periastron $\sqrt{e}\cos \omega$ and $\sqrt{e}\sin \omega$. The priors for RV {offsets $\gamma_{\text{inst}}$} and jitters $s_{\text{inst}}$ were initialized as free parameters for each instrument, set to the mean and standard deviation of the respective RV time series. The posterior orbital parameters and their associated uncertainties were determined from the MAP fit through \texttt{MCMC} sampling of the best-fit Keplerian orbit. The derived orbital parameters and priors are represented in Table \ref{tab: orbital_param}. For the 1-planet model, the Keplerian orbital fit is shown in Figure \ref{fig: 1PL_Kep_fit}, with the corresponding posterior parameter plot in Appendix \ref{sec: appendix_subfig_tab}.

\subsection{Planet model - 2~Dra}
\label{subsec: planet}
The RV time series exhibited a peak-to-peak amplitude of approximately $\sim60\,\mathrm{m}\,\mathrm{s}^{-1}$. We incorporated the 340-d period into our 1-planet model in the Kepler orbit fits. In the 1-planet scenario, the posterior orbital parameters were determined as follows: period $P = 340.60^{+1.38}_{-1.33} \text{ d}$, RV semi-amplitude $K = 15.84^{+1.66}_{-1.68}\, \mathrm{m}\,\mathrm{s}^{-1}$, and eccentricity $e = 0.06^{+0.16}_{-0.02}$. Using the posterior stellar mass $M_{\star} = 1.12^{+0.03}_{-0.04}\,M_{\odot}$ derived from stellar tracks (Section \ref{sec: stellar_model}), we calculated a minimum planet mass of $M_{P}\sin i = 0.59^{+0.06}_{-0.06}\,M_{\text{Jup}}$ and semi-major axis $a = 0.991^{+0.003}_{-0.003} \text{ AU}$ for 2~Dra b. The single-planet scenario exhibits strong statistical significance over the no-planet scenario ($\Delta\text{BIC} \simeq 40$), supporting potential RV variability. 

The RV scatters in the no-planet model ($\sim18$--$22 \, \mathrm{m}\,\mathrm{s}^{-1}$) and the rms of RV residuals, both with and without planet(s) ($\sim16$--$20 \, \mathrm{m}\,\mathrm{s}^{-1}$), exceed the predicted oscillation velocity amplitude\footnote{$K_{\text{osc}} = \frac{L/L_{\odot}}{M/M_{\odot}} \cdot (23.14 \pm 1.4)\, \mathrm{cm}\,\mathrm{s}^{-1}$ \citep{kjeldsen1994amplitudes}} of $K_{\text{osc}} = 12.09 \pm 0.84\, \mathrm{m}\,\mathrm{s}^{-1}$. This indicates that $p$-mode oscillations {(along with other variabilities previously discussed in Section \ref{sec: determine_nature_signal})} are unlikely to be the dominant sources of variation, though additional variations persist. Neither \textit{TESS}, Hipparcos, nor ASAS-SN photometry revealed significant brightness variations associated with 2~Dra's RV variations.

However, given the predicted astrometric perturbations for $\sim0.5 M_{\text{Jup}}$ around a solar-mass star, the 340-d candidate still provides negligible astrometric signal $(\alpha \simeq 5.93 \, \mu \text{as})$
\footnote{
$\alpha [\mu\text{as}] = \frac{M_{P}\sin i}{M_{*}}(\frac{a}{d})$, where $\alpha$ (in unit of $\mu\text{as}$) is astrometric perturbation,} at $\sim1 \text{AU}$. This weak astrometry signal is substantially well below detection thresholds compared to \textit{Gaia} precisions for bright stars (e.g., $\sim$ many tens to 100 $\, \mu \text{as}$). This finding therefore indicates inadequate evidence for the astrometric detection for a 340-d candidate. 

If assuming the 340-d RV signal is a hypothetical planet, the supported arguments {but with a strong caveat (see Discussion)} for a planet hypothesis are:  
\begin{enumerate}
\item Clear, stable, and regular RV variation, like Keplerian motion, is consistently periodic in phase, period, and {amplitude}. 
\item RV does not {significantly} correlate with any indicators, indicating the RV is not from any known activity. 
\end{enumerate}
However, these could not provide sufficient evidence to {completely} refute no 340-d planet hypothesis because
\begin{enumerate}
\item unknown stellar activity cannot be currently fully diagnosed at 340 days, as previously discussed, {and}
\item 
{localized RV phase/amplitude modulations are apparent at specific timespans.
}
\end{enumerate}
{While both issues are not fully explained, we cannot entirely exclude subtle stellar contributions for the 340-d RV signal.}

\section{Discussion \& Summary} 
\label{sec: discussion}
Although 2~Dra is a relatively quiet star, its year-long variability remains substantially noisy. To disentangle these mixed variation sources, we conducted both correlation and phase-stability analyses. Our examination of the 340-d RV signal in 2~Dra confirms that it does not arise from {IP} drift or from yearly sampling aliases near this period. This conclusion is supported by several factors: The RV phases remain largely stable (unlike the fluctuating IP phases), and we find {no significant IP–RV correlations} or {WF–RV correlations} ($|r|\lesssim 0.1$). 
By contrast, the 1540-d RV variation exhibits a sampling signature ($|r|={0.5}$),  which rules out the outer candidate. 
Potential stellar sources, including rotation {($270$–$320$\,d)}, spots, $p$-mode oscillations {($\sim 0.36$\,d, $12~\mathrm{m\,s^{-1}}$)}, and chromospheric activity, show only weak correlations to the RVs ($|r|\lesssim {0.33}$). Thus, these cannot account for the 340-d RV signal {($\simeq 15~\mathrm{m\,s^{-1}}$)}, due to mismatches in period and/or amplitude. However, long-period stellar variations associated with the 340-d RV signal warrants careful consideration from different perspectives.

In Long-Period Variable (LPV) stars,  \citet{suresh2024automated} identified numerous LPV {pulsations ($\sim200 \text{-}500 \text{ d}$)}, particularly \textit{Miras} in ARGB giants, mainly in $\sim$340-d periods. Although this period is common among LPV variables, such variability is unlikely to explain the signal in 2~Dra. First, 2~Dra (an RGB) is not sufficiently luminous to be capable of sustaining LPV modulation ($L < 10^{3}$--$10^{4}L_{\odot}$). Second, the RV amplitude of 2~Dra is $\sim$1000 times smaller than the characteristic LPV pulsation amplitude. LPV pulsations are therefore not a significant contributor to the 340-d signal observed in 2~Dra.

In Long Secondary Period (LSP) stars, the LSP variations {($\sim300\textit{-}1500 \text{ d}$)} may arise from ellipsoidal binaries {or} dust cloud eclipses \citep{soszynski2014light, soszynski2021binarity}. In the binary scenario, $\epsilon$ Cyg A, a K0III giant with $1M_{\odot}$, $57L_{\odot}$, $11R_{\odot}$, and $\nu_{\text{max}}\simeq 32 \ \mu\text{Hz}$, exhibited a nearly 300-d RV variation \citep{heeren2021precise} and shares similar characteristics with 2~Dra. However, unlike $\epsilon$ Cyg, 2~Dra shows no evidence of a stellar or sub-stellar companion. For a hypothetical 340-d orbiting companion (1 AU, $0.5M_{\text{Jup}}$), the resulting RV amplitude would be too weak to {be produced by} an eclipsing LSP binary through tidal distortion beyond the Roche lobe ($K \ll 1$--$10 \, \mathrm{km}\,\mathrm{s}^{-1}$). Although the occulting $\sim$340-d dust cloud {around} LSP stars sounds tempting, non-sinusoidal light curve and RV curve could have been asymmetric by their tidal interactions from a dust cloud or a forming companion over time \citep{soszynski2014light}. From our observations, the 340-d variation is relatively symmetric with $\lesssim60 \, \mathrm{m}\,\mathrm{s}^{-1}$ peak-to-peak RV amplitude, and maintains a {largely} consistent phases over time. Consequently, the LSP binarity model is barely suited to 2~Dra by the extrinsic variation.

However, the stellar origin of 2~Dra's 340-d RV variation {may indicate} a {small-amplitude} intrinsic pulsation, particularly oscillatory convective modes ({OCMs} \cite{saio2015oscillatory, spaeth2024non}). This is inferred from the {similar phase/amplitude modulation behaviors} of the 340-d RV and 270-d $\text{BIS}'$ signals. As demonstrated in several cases, K-giants, e.g., $\alpha$~Tau, $\gamma$~Dra, and 42~Dra initially passed the planet tests (showing no bisector-activity correlations and no phase/amplitude drifts in early observations). However, later long-term monitoring and multi-band observations revealed strong amplitude/phase modulations or multi-periodic behavior consistent with intrinsic stellar oscillations (OCMs), refuting the planet candidates. This insight underscores that passing the planet test is critical and must be persistent in the long-term observations in K-giant domain \citep{hatzes2015long, hatzes2018radial, reichert2019precise, hatzes2025no}. \citet{Dollinger2021sanity} noted further, for the large-radii giant hosts with $R_{*} \geq 21R_\odot$. Many reported $\sim$300-800 d planet candidates, instead, likely pointed to a stellar origin (e.g., rotation-modulated structure or non-radial oscillations), mostly beyond the $20R_\odot$ giant stars (e.g., 42~Dra with $\sim20R_\odot$). Though 2~Dra (at $10R_{\odot}$) is well below this large-radius regime ($<21R_\odot$) and may not be subject to this radius-dependent mechanism, the localized phase/amplitude RV modulations in 2~Dra may arguably be involved in these cases. This may point to another phenomenon for the RV variations. Regardless, without additional photometric or multi-band (optical + NIR) evidence, we cannot conclusively exclude intrinsic long-period oscillations (e.g., OCM).

The RV variation here may also plausibly result from mixed subtle variations caused by unexplored stellar activity from one or multiple sources. Consequently, our understanding of giant star mechanisms needs further expansion, particularly in the small-amplitude, low-luminosity regime (i.e., variations of $\sim$tens of $\mathrm{m}\,\mathrm{s^{-1}}$ at $L<100L_{\odot}$) for quiet slow-rotating stars. The pulsation model by \citet{spaeth2024non} provides a promising tool for simulating {OCMs} in {a highly} luminous {cluster} giant, successfully tracking {pulsation} cycles with amplitudes of $\sim$0.1-1 $\mathrm{km}\,\mathrm{s}^{-1}$ in both RV and stellar indicators. However, for our giant {stars (RGB/CHeB)} in the low-luminosity regime, the small pulsation amplitudes ($\sim$tens of $\mathrm{m}\,\mathrm{s}^{-1}$) make it challenging to detect or model the {pulsation} phase cycles in RV and activity indicators. Although 2~Dra's RV amplitude differs from typical pulsation amplitudes, {OCMs} cannot be entirely excluded, even in the low-luminosity regime. When theoretical and observational gaps are addressed, the nature of 2~Dra's 340-d variation will become clearer, emphasizing the need for continuous monitoring {in multi-band observations (e.g., NIR)} and refined stellar models for this type of evolved star. Since stellar variations in 2~Dra cannot be completely ruled out, and the RV signal exhibits {phase-}amplitude drifts that are {apparently} non-periodic, the existence of a hypothetical companion with a 340-d period remains uncertain. We emphasize that any hypothetical orbital solution could be readily refuted once 2~Dra's intrinsic variations are better understood, due to the characteristics of GK giants.

\enlargethispage{\baselineskip}

\begin{ack}
This study is supported by the National Natural Science Foundation of China under grant No. 12588202, and the National Key R\&D Program of China No. 2024YFA1611903. 
U.S. would like to gratefully acknowledge financial support from the Alliance of National and International Science Organizations (CAS-ANSO) scholarship under grant No. 2023ANP0171 by the Chinese Academy of Sciences (CAS) and National Astronomical Observatories of China (NAOC). H.Y.T. appreciates the support by the EACOA/EAO Fellowship Program under the umbrella of the East Asia Core Observatories Association. M.H. acknowledges support from NASA grant 80NSSC24K0228. This study is supported by National Natural Science Foundation of China grants 62127901, National Key R\&D Program of China No. 2025YFE0102100, 2024YFA1611802 and 2025YFE0213204, the National Astronomical Observatories Chinese Academy of Sciences No. E4TQ2101, the China Manned Space Project with NO. CMS-CSST-2025-A16 and the Pre-research project on Civil Aerospace Technologies No. D010301 funded by China National Space Administration (CNSA). This research was in part supported by JSPS KAKENHI Grant Numbers JP23244038, JP16H02169. We would like to give a huge thank you to the Okayama Planet Search Program (OPSP)'s team for providing access to the EAPS-Net observations from the HIDES instrument with the support of the researchers at Okayama Astrophysical Observatory (OAO) for providing data acquisition. We specifically thank all anonymous referees for providing the constructive and valuable feedback on this manuscript. 
\end{ack}

\begin{appendix}

\section{Software}
\texttt{astropy} \citep{robitaille2013astropy}, 
\texttt{isoclassify} \citep{huber2017asteroseismology,berger2020gaia}, \texttt{MIST} \citep{choi2016mesa}, \texttt{pySYD} \citep{Chontos2022pySYD}, \texttt{radvel} \citep{fulton2018radvel}, \texttt{sbgls} \citep{mortier2017stacked}

\section{Supplementary Figures \& Tables}
\label{sec: appendix_subfig_tab}
\begin{figure*}[!ht]
    \centering
    \includegraphics[width=1.\linewidth]{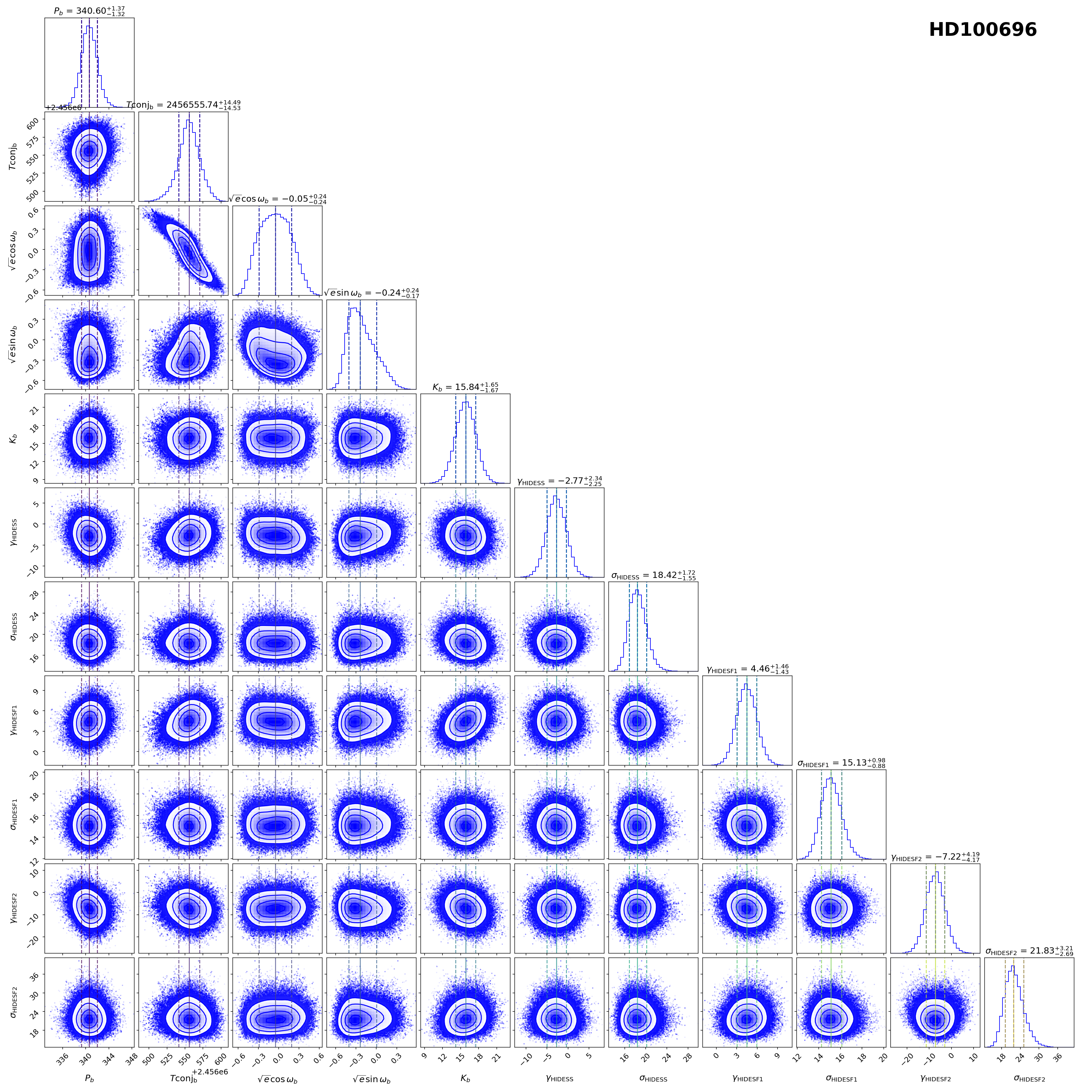}
    \caption{Corner plot of the maximized-likelihood orbital parameters in the 1-planet model.
    }
    \label{fig: 1PL_mcmc_param}
\end{figure*}

\begin{figure*}
    \centering
    \includegraphics[width=1.\linewidth]{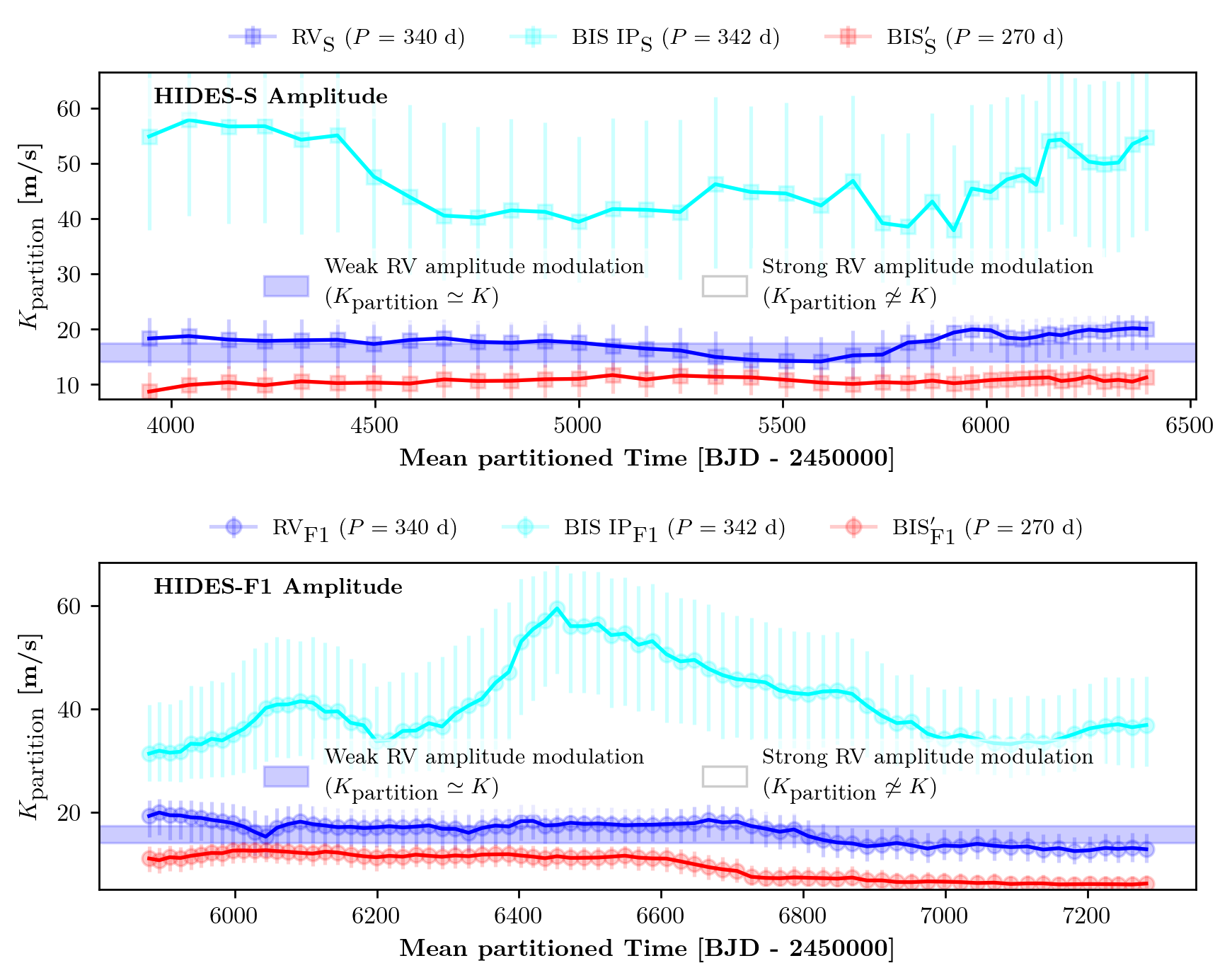}
    \caption{
    Amplitude partitions of 2~Dra signals. The modulation stability is indicated by {relative flatness} across partitions in the \textit{weak modulation zones} between HIDES-S (\textit{top panel}) and HIDES-F1 \textit{(bottom panel}) measurements, 
    as similarly detailed in Figure \ref{fig: phase_stability_2Dra_SS_F1}. 
    Blue region is the posterior $K$ with $1\sigma$ errorband of full RV data from Table \ref{tab: orbital_param} (1-planet model),
    suggesting the \textit{weak RV amplitude modulation zone} for the RV's partitioned $K$. 
    When partitioned $K$ is fully outside of the blue region; significant RV amplitude modulations may be implied.
    }
    \label{fig: amp_stability_2Dra_SS_F1}
\end{figure*}

Table \ref{tab: photometric_param} shows the photometry parameters of 2~Dra used in our work. 
The best-fitted \texttt{MCMC} sampling of 1-planet model parameters is in Figure \ref{fig: 1PL_mcmc_param}.
Amplitude periodic/stability test of 2~Dra's RV, BIS$'$, BIS IP variations between HIDES-S and -F1 is illustrated in Figure \ref{fig: amp_stability_2Dra_SS_F1}.

\begin{table}[!htbp]
\caption{Photometric parameters of 2~Dra}
\begin{tabularx}{0.5\textwidth}{lc} 
\hline\hline
Parameters & Values \\
\hline\hline
$B$-band magnitude $B$ & 6.425 $\pm$ 0.003$^{h}$ \\
$G$-band magnitude $G$ & $4.920 \pm 0.003^{\gamma}$ \\
Blue-photometer $G$-band mag $G_{BP}$ & $5.417^{\gamma}$ \\ 
Red-photometer $G$-band mag $G_{RP}$ & $4.265^{\gamma}$ \\ 
$R$-band magnitude $R$ & $4.6^{u}$ \\
$J$-band magnitude $J$ & $3.647 \pm 0.294^{\mu}$ \\ 
$H$-band magnitude $H$ & $3.132 \pm 0.228^{\mu}$ \\ 
$K$-band magnitude $K_s$ & $2.881 \pm 0.304^{\mu}$ \\ 
Interstellar extinction $A_{V}$ [mag] & ${0.022^{+0.044}_{-0.042}}^{\dagger,g}$ \\ 
Absolute magnitude $M_{V}$ & ${0.657^{+0.030}_{-0.030}}^{\dagger,b}$ \\
Bolometric magnitude $M_{\rm bol}$ & ${0.381^{+0.031}_{-0.030}}^{\dagger,b}$\\
\hline
\end{tabularx}
\vspace{1mm}
{\footnotesize \raggedright 
Note: $^{\dagger}$ = Evolutionary Track (This work). $^{g}$ = Grid model. \\ 
$^{b}$ = Bolometric correction. \\
Ref: $^{\gamma}$ = Gaia EDR3  \citep{brown2021gaia}. \\
$^{h}$ = Hipparcos \citep{esa1997vizier}. \\
$^{u}$ = USNO-B \citep{monet2003usno}. \\ 
$^{\mu}$ = 2MASS \citep{cutri20032mass}. \par
}
\label{tab: photometric_param}
\end{table}

\end{appendix}

\enlargethispage{\baselineskip}


\end{document}